\documentclass[3p,preprint,hyperref]{elsarticle}
\input{prolegomena}
\begin{document}

\null\vskip-60pt \hfill
\begin{minipage}[t]{4cm}
FR-PHENO-2016-004 \\
\end{minipage}

\begin{frontmatter}
\title{Low energy behaviour of standard model extensions}

\author[freiburg]{Michele Boggia\fnref{support}}
\ead{michele.boggia@physik.uni-freiburg.de}

\author[torino]{Raquel Gomez-Ambrosio\fnref{support}}
\ead{raquel.gomez@to.infn.it}

\author[torino]{Giampiero Passarino\fnref{support}}
\ead{giampiero@to.infn.it}

\address[freiburg]{\csumb}

\address[torino]{\csuma}

\fntext[support]{\support}

\begin{abstract}
 \noindent
The integration of heavy scalar fields is discussed in a class of BSM models, containing 
more that one representation for scalars and with mixing. The interplay between integrating out 
heavy scalars and the Standard Model decoupling limit is examined. In general, the latter cannot 
be obtained in terms of only one large scale and can only be achieved by imposing further 
assumptions on the couplings. Systematic low-energy expansions are derived in the more general, 
non-decoupling scenario, including mixed tree-loop and mixed heavy-light generated operators. 
The number of local operators is larger than the one usually reported in the literature.
\end{abstract}
\begin{keyword}
Perturbative calculations,
Extensions of electroweak Higgs sector,
Effective Field Theory
\PACS 12.38.Bx, 12.60.Fr, 14.80.Ec, 12.60.-i  
\end{keyword}

\end{frontmatter}


\section{Introduction \label{Intro}}
There are two ways to use effective field theories (EFT), the bottom-up approach and the 
top-down approach.
To apply the first, we must distinguish between two scenarios: a) there is no relevant theory at the energy scale under consideration, in which case one has 
to construct a Lagrangian from the symmetries that are relevant at that scale, b) there is already some EFT, \eg Standard Model (SM) EFT or SMEFT, which represents 
the physics in a region characterized by a cut-off parameter $\Lambda$. At higher energies, 
new phenomena might show up and our EFT does not account for them.

In the top-down approach there is some theory, assumed to be ultraviolet (UV) complete 
or valid on a given high energy scale (\eg some BSM model), and the aim is to implement a 
systematic procedure for getting the low-energy theory. A typical example would be the 
Euler-Heisenberg Lagrangian. Systematic low-energy expansions are able to obtain low-energy 
footprints of the high energy regime of the theory.

In the top-down approach the heavy fields are integrated out of the underlying high-energy theory
and the resulting effective action is then expanded in a series of local operator terms.
The bottom-up approach is constructed by completely removing the heavy fields, as opposed to
integrating them out; this removal is compensated by including any new nonrenormalizable 
interaction that may be required. If the UV theory is known, appropriate matching calculations
will follow.  

In this work we will discuss the integration of heavy fields in a wide class of BSM models,
containing more that one representation for scalars, with the presence of mixing. For early
work on the subject, see \Brefs{deBlas:2014mba,Chiang:2015ura}.
One problem in dealing with BSM models is the absence of a well-defined hierarchy of scales,
see~\Bref{Brehmer:2015rna,Biekotter:2016ecg} for a discussion. A second problem, as observed in
\Bref{delAguila:2016zcb}, is that there are cases where the so-called covariant derivative 
expansion~\cite{Gaillard:1985uh,Cheyette:1987qz,Henning:2014wua,Drozd:2015rsp} (CDE) does not 
reproduce all the local operators in the low-energy sector. 

In most BSM models, loop effects are certainly suppressed and the leading observable consequences 
are those generated at tree level. However, considering projections for the precision to be 
reached in LHC Run-II analysis, LO results for interpretations of the data are challenged by 
consistency concerns and are not sufficient, if the cut off scale is in the few TeV range.
Moving to the consistent inclusion of loop effects adds complexity but robustly accommodates 
the precision projected to be achieved in Run-II analyses.

The aim of this paper is not to guess which is the UV completion of the SM chosen by nature, 
but rather to present in a systematic way how the calculation of a (top-down) EFT for 
any realistic model should be done.

The paper is organized as follows: in Sect.~\ref{Gform} we present the general formalism.
The low energy behavior for the singlet extension of the SM is discussed in Sect.~\ref{EFTS} and
the THDM models on Sect.~\ref{EFTD}.
\section{General formalism \label{Gform}}
The most general Lagrangian that we have in mind contains, after mixing, $n$ heavy scalar fields
(charged or neutral) and can be written as
\bq
\Lag_{\BSM} = \Lag_{\mySM} + \Delta \Lag^{(4)} + \Lag^{(4)}_{\mrH} \spc
\quad
\Lag^{(4)}_{\mrH} = \sum_{i_1=0}^{h}\,\cdots\,\sum_{i_k=0}^{\ssI_{k-1}}\,\cdots\,
\sum_{i_{n-1}=0}^{\ssI_{n-2}}\,\mrF^{h}_{i_1\,\dots\,i_n}\,
\PH^{i_1}_1\,\dots\,\PH^{\ssI_{n-1}}_n\;+\;\mbox{h.c.} \spc
\label{genc}
\eq
where $\mrI_k= h - i_1 -\,\dots\,- i_{k-1}$. 
The term $\Lag_{\mySM}$ is the SM Lagrangian and $\Delta \Lag^{(4)}$ contains light fields
only and it is proportional to non-SM couplings (\ie corrections to SM-couplings, due to the 
new interactions). Furthermore, $\mrF^{h}$ is a function of the 
light fields with canonical dimension $4 - h$.

Specific examples for the terms in the Lagrangian of \eqn{genc} are:
$\upPhi^{\dagger}\,\upPhi\,\mrS$, where $\upPhi$ is the standard Higgs doublet and $\mrS$ is 
a singlet; $\upPhi^{\dagger}\,\tau_a\,\upPhi\,\mrT^a$, where $\mrT^a$ is a scalar 
(real or complex) triplet~\cite{SekharChivukula:2007gi,Chen:2008jg};  
$\upPhi^{\dagger}\,\tau_a\,\upPhi^c\,\Xi^{\dagger}\,t^a\,X$
where $\Xi$ is a zero hypercharge real triplet, $X$ a $Y = 1$ complex triplet and
$\upPhi^c$ is the charge conjugate of $\upPhi$, the so-called Georgi-Machacek model,
see~\Bref{Georgi:1985nv}. For a classification of CP even scalars according to their properties 
under custodial symmetry see~\Brefs{Einhorn:1981cy,Low:2012rj}.
For a discussion on fingerprints of non-minimal Higgs sectors, see \Bref{Kanemura:2014bqa}.

There are two sources of deviations with respect to the SM, new couplings and modified
couplings due to VEV mixings, heavy fields. In general, it is not simple to identify
only one scale for new physics (NP); it is relatively simple in the unbroken phase using
weak eigenstates but it becomes more complicated when EWSB is taken into account and one
works with the mass eigenstates. In the second case, one should also take into account
that there are relations among the parameters of the BSM model, typically coupling constants
can be expressed in terms of VEVs and masses; once the heavy scale has been introduced
also these relations should be consistently expanded. Briefly, the SM decoupling limit cannot 
be obtained by making only assumptions about one parameter.
This fact adds additional operators to the SM that are not those caused by integrating 
out the heavy fields.

There are three reasons why published CDE results do not give the full result in 
explicit form (\eg see the $\mcO(\upPhi^3_c)$ terms in Eq.~(2.7) of \Bref{Henning:2014wua}).
\begin{enumerate}
\item The functions $\mrF$ in \eqn{genc} may contain positive powers of the heavy scale, so that 
terms of dimension greater than $2$ in the heavy fields have been retained in our functional 
integral (the linear terms as well). 

\item The second reason is that there are mixed tree-loop--generated operators, see 
Fig.~\ref{MTLrules}, where we show a diagram that, after integration of the internal heavy 
lines and contraction of the external heavy lines gives a contribution $\mcO(\Lambda^{-2})$ 
(here $\mrF_{1,3} \propto \Lambda$).

\begin{figure}[t]
   \centering
   \includegraphics[width=0.8\textwidth, trim = 30 250 50 80, clip=true]{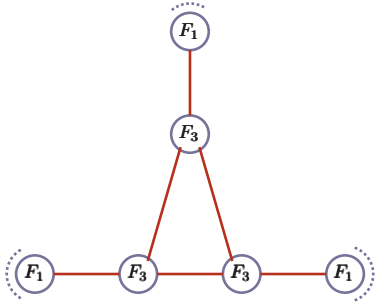}
\vspace{-4.5cm}
\caption[]{Example of mixed tree-loop--generated operator. Solid (red) lines denote heavy fields, 
Blobs denote vertices with additional light lines.}
\label{MTLrules}
\end{figure}

\item The third reason is that there are mixed loops, containing both light and heavy particles. 
\end{enumerate}
In the following we will discuss the full derivation of the low-energy limit for the case of one 
(neutral) heavy field, the generalization being straightforward. Therefore, we write the 
Lagrangian as follows:
\bq
\mcL = \mcL_0 + \PH\,\mcO_2\,\PH + \sum_{n=1}^{4}\,\mcF_n\,\PH^n \spc
\label{Lagtwo}
\eq
where $\mcO_2$ is the Klein-Gordon operator for the heavy field $\PH$. Furthermore, without 
loss of generality, we assume the following behavior
\bqa
\mrF_1(x) = \Lambda\,\mrF_{10}(x) + \Lambda^{-1}\,\mrF_{11}(x) + \Lambda^{-3}\,\mrF_{12}(x) \spc
&\qquad&
\mrF_2(x) = \mrF_{20}(x) + \Lambda^{-2}\,\mrF_{21}(x) + \Lambda^{-4}\,\mrF_{22}(x) \spc
\nl
\mrF_3(x) = \Lambda\,\mrF_{30}(x) + \Lambda^{-1}\,\mrF_{31}(x) + \Lambda^{-3}\,\mrF_{32}(x) \spc
&\qquad&
\mrF_4(x) = \mrF_{40}(x) + \Lambda^{-2}\,\mrF_{41}(x) + \Lambda^{-4}\,\mrF_{42}(x) \spc
\label{Fexp}
\eqa
where $\Lambda$ is the scale controlling the onset of new physics, not necessarily equal
to the mass of the heavy field. The latter, $M_{\PH}$, is expressed in terms of $\Lambda$ by
\bq
M^2_{\PH} = \Lambda^2\,\sum_{n=0}\,\xi_n\,\lpar \frac{\mws}{\Lambda^2} \rpar^n \spc
\label{xipar}
\eq
with coefficients $\xi_i$ that depend on the model. Furthermore, we have truncated the expansion 
at the right level to derive $\mrdim = 6$ operators. 
Finally, $\mrdim\,\mrF_{1n} = \mrdim\,\mrF_{2n} = 2\,(n+1)$ and
$\mrdim\,\mrF_{3n} = \mrdim\,\mrF_{4n} = 2\,n$.

The integration of the heavy mode, $\PH$, gives an effective Lagrangian and results in the 
addition of tree-generated, loop-generated, tree-loop--generated and mixed heavy-light 
loop-generated operators. 
Actually there are two different ways to construct a low-energy theory: one can integrate the
heavy particles by diagrammatic methods, or use functional methods; for both cases see
\Bref{Dittmaier:1995cr}. Our derivation is as follows: consider the functional integral
\bq
\mrW = \int [\mrD\,\PH]\,\exp\Bigl\{ i\,\int d^4 x\,\mcL_{\PH}\Bigr\} \spc
\quad
\mcL_{\PH} = - \frac{1}{2}\,\pdmu\,\PH\,\pdmu\,\PH - \frac{1}{2}\,M^2_{\PH}\,\PH^2 +
             \sum_{n=1}^{4}\,\mrF_n\,\PH^n \spp
\eq
Using standard algorithms we obtain
\bq
\mrW = \exp\Bigl\{i\,\int d^4 y\,
        \sum_{n=2}^{4}\,\mrF_n(y)\,\lpar -\,i\,\delta_{\mrF_1}(y) \rpar^n 
        \Bigr\}\, 
      \int [\mrD\,\PH]\,
      \exp\Bigl\{i\,\int d^4 x\,\mcL^{(0)}_{\PH}\Bigr\} \spc
\label{MTGfunI}
\eq
where we have introduced a free Lagrangian with a source term for the heavy field,
\bq
\mcL^{(0)}_{\PH} = - \frac{1}{2}\,\pdmu\,\PH\,\pdmu\,\PH - \frac{1}{2}\,M^2_{\PH}\,\PH^2 +  
                  \mrF_1\,\PH \spc
\eq
and the functional derivative
\bq
\delta_{\mrF_1}(x) = \frac{\updelta}{\updelta\,\mrF_1(x)} \spp
\eq
It is worth noting that \eqn{MTGfunI} is needed in order to reproduce mixed tree-loop--generated 
operators. Using the well-known result
\bq
\int [\mrD\,\PH]\,\exp\Bigl\{i\,\int d^4 x\,\mcL^{(0)}_{\PH}\Bigr\} = \mrW_0\,
   \exp\Bigl\{ -\,\frac{1}{2}\,\int d^4u\,d^4v\,
      \mrF_1(u)\,\Delta_{\mrF}\lpar u - \mrv \rpar\,\mrF_1(v) \Bigr\} \spc
\eq
where $\mrW_0$ is the $\mrF_1\,$-independent normalization constant and $\Delta_{\mrF}(z)$ is 
the Feynman propagator,
\bq
\Delta_{\mrF}(z)= \frac{1}{(2\,\pi)^4\,i}\,\int d^4 p\,
                  \frac{\exp\{i\,\spro{p}{z}\}}{p^2 + M^2_{\PH} - i\,0} \spp
\eq
The effective Lagrangian (up to order $\Lambda^{-2}$) becomes
\bq
\mrW = \mrW_0\,\exp\Bigl\{ i\,\int d^4 x\,\mcL_{\eff}\Bigr\} \spc            
\qquad
\mcL_{\eff} = \mcL^{\mrT}_{\eff} + \frac{1}{16\,\pi^2}\,\mcL^{\mrL}_{\eff} \spp
\eq
The tree-generated Lagrangian becomes
\bq
\mcL^{\mrT}_{\eff} = \frac{1}{2}\,\xi^{-1}_0\,\mrF^2_{10}
       + \frac{1}{\xi^3_0\,\Lambda^2} \, \Bigl[ \mrF^3_{10}\,\mrF_{30}
       + \frac{1}{2}\,\xi_0\,\lpar 2\,\mrF^2_{10}\,\mrF_{20} - M^2\,\xi_1\,\mrF^2_{10}
          - \pdmu \mrF_{10}\,\pumu \mrF_{10} \rpar
       + \xi^2_0\,\mrF_{10}\,\mrF_{11} \Bigr]
\eq
It is worth noting that there are terms, \eg those proportional to $\mrF_{30}$, that
are left implicit in the published CDE results.

\begin{figure}[t]
   \centering
   \includegraphics[width=0.8\textwidth, trim = 30 250 50 80, clip=true]{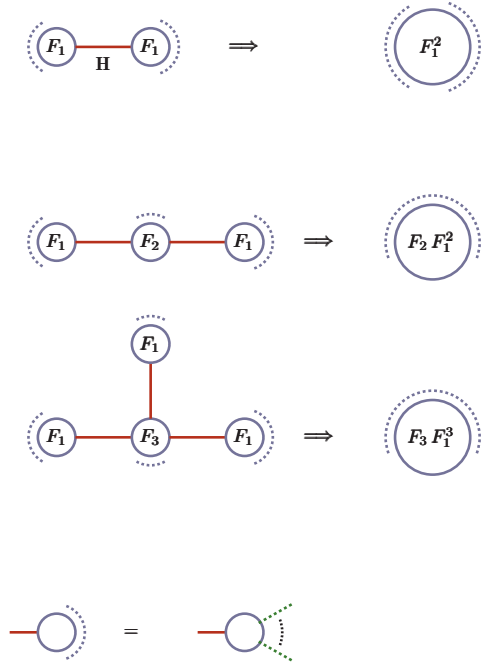}
\vspace{-1.5cm}
\caption[]{Example of tree-generated operators. Solid (red) lines denote heavy fields, dashed
(green) lines denote light fields. Blobs denote vertices with additional light lines.}
\label{TGrules}
\end{figure}

A construction of tree-generated vertices is shown in Fig.~\ref{TGrules} where the result
of functional integration is seen from a different perspective, as a contraction of propagators 
inside diagrams of the full theory.
In \eqn{Fexp} we see why it is not enough to use the Lagrangian truncated at $\mcO(\PH)$ to derive
tree-generated operators; for instance, both $\mrF_1$ and $\mrF_3$ start at $\mcO(\Lambda)$ which 
is enough to compensate the $M^{-6}_{\PH}$ from $\PH$ propagators giving a result at 
$\mcO(\Lambda^{-2})$, as shown in the third row of Fig.~\ref{TGrules}.

\begin{figure}[h]
   \centering
   \includegraphics[width=0.8\textwidth, trim = 30 250 50 80, clip=true]{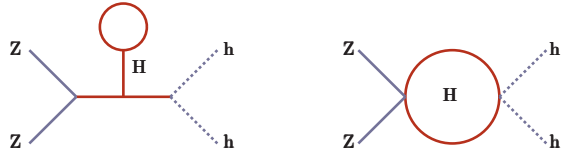}
\vspace{-7.5cm}
\caption[]{Example of loop-generated operators, Solid (red) lines denote the heavy field
$\PH$, solid and dashed (blue) lines denote light SM fields. 
Left figure shows a $\PH$ tadpole to be canceled in the $\beta\,$-scheme while right
figure shows a genuine LG operator.}
\label{TaNT}
\end{figure}

If we restrict to $\mrdim = 6$  operators (i.e to order $\Lambda^{-2}$), loop-induced operators generated by the functional integral belong to three different
cases:
\begin{enumerate}
\item There are triangles with heavy, internal, lines (third row in Fig.~\ref{LGrules}); in the 
limit of large internal (equal) masses, the corresponding loop integral gives 
\bq
\mrC^{\PH}_0 = - \frac{1}{2}\,M^{-2}_{\PH} + \mcO\lpar M^{-4}_{\PH}\rpar \spp
\eq
\item There are also bubbles (first row in Fig.~\ref{LGrules}); in the limit of large internal 
(equal) masses, the corresponding loop integral gives
\bq
\mrB_0 \lpar P^2\,;\,M^2_{\PH}\,,\,M^2_{\PH} \rpar = - \mrB_{00}\lpar M_{\PH} \rpar -
       \frac{1}{6}\,\frac{P^2}{M^2_{\PH}} + \mcO\lpar \frac{P^4}{M^4_{\PH}} \rpar \spc
\label{B00def}
\eq
with $\mrB_{00}(M_{\PH}) = \mrA_0(M_{\PH}) + 1$ and the (dimensionless) one-pont function, 
$\mrA_0(M_{\PH}) \equiv \mrA_0$, is defined in dimensional regularization by
\bq
\muR^{4 - n}\,\int \frac{d^nq}{q^2 + m^2_{\PH}} = i\,\pi^2\,\mrA_0 \,M^2_{\PH}\, =
\,i\,\pi^2\,M^2_{\PH}\,\lpar \frac{2}{n - 4} + \gamma + \ln \pi - 1 + 
\ln\frac{M^2_{\PH}}{\muRs} \rpar \spc 
\label{A0def}
\eq
where $n$ is the space-time dimension, $\gamma$ is the Euler-Mascheroni constant and $\muR$ is 
the renormalization scale. 

\begin{figure}[t]
   \centering
   \includegraphics[width=0.8\textwidth, trim = 30 250 50 80, clip=true]{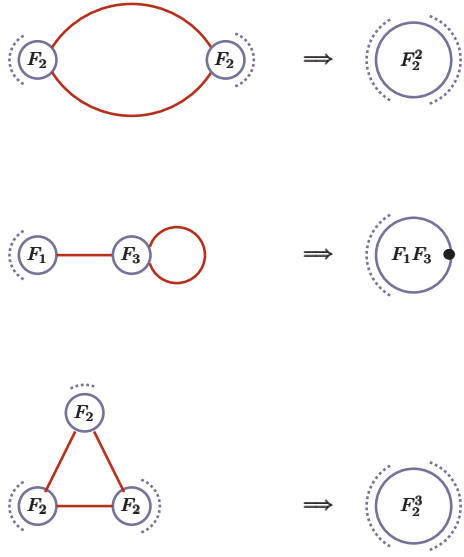}
\vspace{-4.cm}
\caption[]{Example of loop-generated operators. Solid (red) lines denote heavy fields, 
blobs denote vertices with additional light lines. The (black) bullet denotes a tadpole}
\label{LGrules}
\end{figure}

\item Finally, there are tadpoles, as shown in the second row of Fig.~\ref{LGrules}. The 
treatment of tadpoles, \ie their cancellation, is model dependent. Here we present the list of 
tadpoles and postpone discussing their cancellation until Sect.~\ref{Tad}.
Therefore, in writing $\mcL^{\mrL}_{\eff}$ we split the Lagrangian into two parts: the one
containing tadpoles and the one without.
%
Examples are shown in Fig.~\ref{TaNT} for the $\Ph^2\,\PZ^2$ operator: the left diagram is
a $\PH$ tadpole while the right one is a genuine LG operator.
\end{enumerate}

With the $\Lambda$ power counting of \eqn{Fexp} and loop power counting of \eqn{PCLI} it is 
easily seen that boxes of heavy lines start contributing only at $\mcO(\Lambda^{-4})$.

The $\mrF$ functions, defined in \eqn{Fexp}, are polynomials in the light fields of the form
$\mrF_{ij}= \mrF^{\Ph}_{ij} + \mrF^{\rest}_{ij}$, where 
$\mrF^{\Ph}_{ij} = \kappa_{ij}\,\Ph$ and ``rest'' contains two or more fields. 
The result, split in non-tadpole (NT) and tadpole (T) contributions, is as follows 
\begin{center}
\boxed{ 
\mcL^{\mrL}_{\eff\,\mrN\mrT} =
\xi_0\,\Lambda^2\,\mcL^{\mrL\,,\,2}_{\eff\,\mrN\mrT} + 
\frac{1}{\xi^2_0}\,\mcL^{\mrL\,,\,0}_{\eff\,\mrN\mrT} + 
\frac{1}{\xi^4_0\,\Lambda^2}\,\mcL^{\mrL\,,\,- 2}_{\eff\,\mrN\mrT} 
}
\end{center}
\bqa  
\mcL^{\mrL\,,\,2}_{\eff\,\mrN\mrT} &=& \mrA_0\,\mrF^{\rest}_{20} \spc
\nl\nl
\mcL^{\mrL\,,\,0}_{\eff\,\mrN\mrT} &=&  -\,\mrB_{00}\,\lpar
   9\,\mrF^2_{10}\,\mrF^2_{30} + 6\,\xi_0\,\mrF_{10}\,\mrF_{20}\,\mrF_{30} + \xi^2_0\,\mrF^2_{20}
   \rpar 
   + 6\,\mrA_0\,\xi_0\,\mrF^2_{10}\,\mrF_{40} 
   + \xi^2_0\,\mrA_0\,\lpar \xi_1\,M^2\,\mrF^{\rest}_{20} + \xi_0\,\mrF^{\rest}_{21} \rpar \spc 
\nl\nl
\mcL^{\mrL\,,\,- 2}_{\eff\,\mrN\mrT} &=&
         18
         \,\lpar 1 - 3\,\mrB_{00} \rpar
         \,\mrF^2_{10}\,\mrF^2_{30}\,\lpar \mrF_{10}\,\mrF_{30} + \xi_0\,\mrF_{20} \rpar
\nl
{}&+&
         2
         \,\xi_0
         \,\mrB_{00}
         \,\Bigl[
          9\,M^2\,\xi_1\,\mrF^2_{10}\,\mrF^2_{30} 
        - 3\,\xi_0\,\mrF_{10}\,\mrF_{30}\,\lpar 3\,\mrF_{10}\,\mrF_{31} 
        + 3\,\mrF_{11}\,\mrF_{30} - M^2\,\xi_1\,\mrF_{20} \rpar 
\nl
{}&-& 3\,\xi^2_0\,\lpar \mrF_{10}\,\mrF_{21}\,\mrF_{30} + 
                        \mrF_{10}\,\mrF_{20}\,\mrF_{31} + 
                        \mrF_{11}\,\mrF_{20}\,\mrF_{30} \rpar 
        - \xi^3_0\,\mrF_{20}\,\mrF_{21}
            \Bigr]
\nl
{}&+&
         6
         \,\xi^2_0
         \,\lpar 1 - 2\,\mrB_{00} \rpar
         \,\mrF_{10}\,\mrF^2_{20}\,\mrF_{30}
\nl
{}&-&
         \frac{1}{6}
         \,\xi_0
         \,\Bigl[
          9\,\mrF^2_{30}\,\pdmu \mrF_{10}\,\pumu \mrF_{10} 
        + 6\,\xi_0\,\mrF_{30}\,\pdmu \mrF_{10}\,\pumu \mrF_{20} 
          - \xi^2_0\,\lpar 4\,\mrF^3_{20} - \pdmu \mrF_{20}\,\pumu \mrF_{20} \rpar
            \Bigr] 
\nl
{}&-& 6\,\mrA_0\,\xi^2_0
         \,\mrF_{10}\,\Bigl[ 
     M^2\,\xi_1\,\mrF_{10}\,\mrF_{40} - \xi_0\,\lpar 
      \mrF_{10}\,\mrF_{41} + 2\,\mrF_{11}\,\mrF_{40} \rpar \Bigr] 
%
\nl
{}&-& 12\,\xi_0\,\mrB_{00}\,\mrF_{40}\,\mrF^2_{10}\,
         \lpar 3\,\mrF_{10}\,\mrF_{30} +\xi_0\,\mrF_{20} \rpar 
         + \xi^4_0\,\mrA_0\,\Bigl[
           M^2\,\lpar \xi_2\,M^2\,\mrF^{\rest}_{20} + \xi_1\,\mrF^{\rest}_{21} \rpar
         + \xi_0\,\mrF^{\rest}_{22} \Bigr] \spc
\label{mbasisa}
\eqa
\begin{center}
\boxed{
\mcL^{\mrL}_{\eff\,\mrT} = \mrA_0\,\Bigl[
\Lambda^2\,\mcL^{\mrL\,,\,2}_{\eff\,\mrT} + 
\frac{1}{\xi^2_0}\,\mcL^{\mrL\,,\,0}_{\eff\,\mrT} + 
\frac{1}{\xi^4_0\,\Lambda^2}\,\mcL^{\mrL\,,\,- 2}_{\eff\,\mrT} \Bigr] 
}
\end{center}
\bqa
\mcL^{\mrL\,,\,2}_{\eff\,\mrT} &=&  3\,\mrF_{10}\,\mrF_{30} + \xi_0\,\mrF^{\Ph}_{20}  \spc
\nl\nl
\mcL^{\mrL\,,\,0}_{\eff\,\mrT} &=& 
        9\,\mrF^2_{10}\,\mrF^2_{30} + 
        6\,\xi_0\,\mrF_{10}\,\mrF_{20}\,\mrF_{30} + 
        \xi^2_0\,\lpar 3\,\mrF_{10}\,\mrF_{31} + 3\,\mrF_{11}\,\mrF_{30} + 
        M^2\,\xi_1\,\mrF^{\Ph}_{20} \rpar + \xi^3_0\,\mrF^{\Ph}_{21} \spc
\nl\nl
\mcL^{\mrL\,,\,- 2}_{\eff\,\mrT} &=& 
          54\,\mrF^3_{10}\,\mrF^3_{30} 
        - 18\,\xi_0\,\mrF^2_{10}\,\mrF^2_{30}\,\lpar M^2\,\xi_1 - 3\,\mrF_{20} \rpar 
\nl
{}&+& 6\,\xi^2_0\,\Bigl[ \mrF_{10}\,\lpar 3\,\mrF_{10}\,\mrF_{30}\,\mrF_{31} + 
               3\,\mrF_{11}\,\mrF^2_{30} - M^2\,\xi_1\,\mrF_{20}\,\mrF_{30} + 
               2\,\mrF^2_{20}\,\mrF_{30} \rpar - 
               \mrF_{30}\,\pdmu \mrF_{10}\,\pumu \mrF_{20}
             \Bigr]
\nl
{}&+& 3\,\xi^3_0\,\Bigl[ 2\,\mrF_{10}\,\lpar \mrF_{21}\,\mrF_{30} + 
                          \mrF_{20}\,\mrF_{31} \rpar + 
                       2\,\mrF_{11}\,\mrF_{20}\,\mrF_{30} - 
                          \pdmu \mrF_{10}\,\pumu \mrF_{31}
                \Bigr] 
\nl
{}&+& \xi^4_0\,\lpar 3\,\mrF_{10}\,\mrF_{32} + 3\,\mrF_{11}\,\mrF_{31} + 
                         3\,\mrF_{12}\,\mrF_{30} + 
                         M^4\,\xi_2\,\mrF^{\Ph}_{20} + 
                         M^2\,\xi_1\,\mrF^{\Ph}_{21} \rpar 
         + \xi^5_0\,\mrF^{\Ph}_{22} \spp
\label{mbasisb}
\eqa
We are still missing mixed loop contributions; they are clearly model dependent and will be
discussed in details in Sect.~\ref{EFTS}. The results of \eqns{mbasisa}{mbasisb} form a basis 
of local operators.
\section{Low energy behavior for the singlet extension of the SM \label{EFTS}}
The SM scalar field $\Upphi$ (with hypercharge $1/2$) is defined by
\bq
\Upphi = \frac{1}{\srt}\,\left(
\begin{array}{c}
\Pht + \srt\,\mrv + i\,\Ppz \\
\srt\,i\,\Ppm
\end{array}
\right) \, ,
\eq
$\Pht$ is the custodial singlet in $\lpar 2_{\ssL}\,\otimes\,2_{\ssR}\rpar = 1\,\oplus\,3$.
Charge conjugation gives $\upPhi^c_i = \ep_{ij}\,\upPhi^*_j$.
\subsection{Notations and conventions \label{NandC}}
The Lagrangian giving the singlet 
extension~\cite{Silveira:1985rk,Schabinger:2005ei,Pruna:2013bma,Robens:2015gla,Robens:2016xkb} 
of the SM (SESM) is
\bq
\mcL= - \lpar \mrD_{\mu}\,\Upphi \rpar^{\dagger}\,\mrD_{\mu}\,\Upphi
      - \pdmu\,\upchi\,\pdmu\,\upchi
      - \mu^2_2\,\Upphi^{\dagger}\,\Upphi
      - \mu^2_1\,\upchi^2
      - \frac{1}{2}\,\lambda_2\,\lpar \Upphi^{\dagger}\,\Upphi \rpar^2
      - \frac{1}{2}\,\lambda_1\,\upchi^4
      - \lambda_{12}\,\upchi^2\,\Upphi^{\dagger}\,\Upphi \spc
\label{Lagone}
\eq
where the singlet field and the covariant derivative $\mrD_{\mu}$ are 
\bq
\upchi = \frac{1}{\srt}\,\lpar \Pho + \mrv_{\mrs} \rpar \spc
\quad
\mrD_{\mu} = \pdmu - \frac{i}{2}\,g\,B^a_{\mu}\,\uptau_a - \frac{i}{2}\,g\,g_{_1}\,B^0_{\mu} \spc
\label{covD}
\eq
with $g_{_1} = -\stW/\ctW$ and where $\uptau^a$ are Pauli matrices while $\stW(\ctW)$ is the
sine(cosine) of the weak-mixing angle. Furthermore
\bq
\PWpmmu = \frac{1}{\srt}\,\lpar B^1_{\mu} \mp i\,B^2_{\mu}\rpar \spc
\qquad
\PZmu = \ctW\,B^3_{\mu} - \stW\,B^0_{\mu} \spc
\quad
\PA_{\mu} = \stW\,B^3_{\mu} + \ctW\,B^0_{\mu} \spc
\eq
\bq
F^a_{\mu\nu} = \pdmu\,B^a_{\nu} - \pdnu\,B^a_{\mu}
+ g_{_0}\,\epsilon^{a b c}\,B^b_{\mu}\,B^c_{\nu} \spc
\quad
F^0_{\mu\nu} = \pdmu\,B^0_{\nu} - \pdnu\,B^0_{\mu} \spp
\eq
Here $a,b,\dots = 1,\dots,3$. We define the $\PW$ (bare) mass and a new mass, $M_{\mrs}$, which will play the role of cut-off scale $\Lambda$,
\bq
M^2 = \frac{1}{2}\,g^2\,\mrF^2 \spc
\qquad
M^2_{\mrs} = \frac{1}{4}\,g^2\,\mrv^2_{\mrs} \spp
\eq
In order to write \eqn{Lagone} in terms of mass eigenstates we introduce
\bq
R^2 = \lpar \lambda_2\,\mrv^2 - \frac{1}{2}\,\lambda_2\,\mrv^2_{\mrs} \rpar ^2 +
      2\,\lpar \lambda_{12}\,\mrv\,\mrv_{\mrs} \rpar^2 \spp
\eq
The mixing angle is defined by
\bq
\Ph = \ca\,\Pht - \sa\,\Pho \spc
\qquad
\PH = \sa\,\Pht + \ca\,\Pho \spc
\eq
\bq
\sin(2\,\alpha) = \srt\,\lambda_{12}\,\mrv\,\mrv_{\mrs}\,R^{-1} \spc
\qquad
\cos(2\,\alpha) = \lpar - \lambda_2\,\mrv^2 + 
                  \frac{1}{2}\,\lambda_1\,\mrv^2_{\mrs} \rpar\,R^{-1} \spp
\eq
Nex, we can eliminate $\mu_1, \mu_2$ in \eqn{Lagone},
\bq
\mu^2_2 = - 2\,\frac{\lambda_2}{g^2}\,M^2 - 2\,\frac{\lambda_{12}}{g^2}\,M^2_{\mrs} \spc
\qquad 
\mu^2_1 = - 2\,\frac{\lambda_1}{g^2}\,M^2_{\mrs} - 2\,\frac{\lambda_{12}}{g^2}\,M^2 \spp
\label{quadmass}
\eq
We keep $\lambda_1$ and $\lambda_{12}$ as free parameters and take the limit $M_{\mrs} \to \infty$.
Following \Bref{Gorbahn:2015gxa} we will assume that the ratio of couplings is of the
order of a perturbative coupling, \ie $\lambda_{12}/\lambda^2_1 < 1/2$.
First we eliminate $\lambda_2$,
\bq
\lambda_2 = \frac{1}{4}\,g^2\,\frac{m^2_{\Ph}}{M^2} + g^2\,\frac{t^2_3}{t_1} +
\frac{1}{4}\,g^2\,\frac{t^2_3}{t^2_1}\,\frac{m^2_{\Ph}}{M^2_{\mrs}} +
\mcO\lpar M^{-4}_{\mrs} \rpar \spc
\eq
where $\lambda_{1} = t_1\,g^2$ and $\lambda_{12} = t_3\,g^2$. Similarly, we obtain
the expansion for $\sin (\alpha) $ and $ \cos(\alpha)$  ($\sa, \ca$),
\bq
\ca = 1 - \frac{1}{2}\,\frac{t^2_3}{t^2_1}\,\frac{M^2}{M^2_{\mrs}} + 
        \mcO\lpar M^{-4}_{\mrs} \rpar \spc
\quad
\sa = \frac{t_3}{t_1}\,\frac{M}{M_{\mrs}}\,\Bigl[ 1 +
 \lpar \frac{t_2}{t_1} - \frac{3}{2}\,\frac{t^2_3}{t^2_1}\rpar\,\frac{M^2}{M^2_{\mrs}} 
  \Bigr] + \mcO\lpar M^{-5}_{\mrs} \rpar  \spp
\eq
\begin{remark}
The behavior of $\sa$ is not selected a priori but follows from the hierarchy of VEVs.
Additional suppression of the heavy mode can be imposed by requiring
$\lambda_{12} \propto g^2\,M/M_{\mrs}$, \ie this additional suppression of $\sa$ is an 
independent condition. In any case, the SM decoupling limit cannot be obtained by making only 
assumptions about one parameter. We adopt the more conservative approach, considering the 
non-decoupling limit and $\lambda_{12}$ as a free parameter of the effective theory.
\end{remark}
Finally, the relation between $M_{\PH}$ and $M_{\mrs}$ is
\bq
M^2_{\PH} = 4\,t_1\,M^2_{\mrs}\,\Bigl[ 1 + \frac{t^2_3}{t^2_1}\,\frac{M^2}{M^2_{\mrs}} +
            \mcO\lpar M^{-4}_{\mrs} \rpar \Bigr] \spc
\eq
where $M_{\PH}$ is the mass of the heavy Higgs boson and $\lambda_1 = t_1\,g^2$.
The $\xi$ parameters of \eqn{xipar} are defined by
\bq
\xi_0 = 4\,t_1, \qquad
\xi_1= 4\,\frac{t^2_3}{t_1}, \qquad
\xi_2= \frac{t^2_3}{t^2_1}\,\frac{M^2_{\Ph}}{M^2}
\label{SExidef}
\eq
Assuming that $M_{\PH} \gg M_{\Ph}$ we construct the corresponding low scale approximation
of the model~\cite{deBlas:2014mba,Chiang:2015ura}. There are three options that will be discussed 
in the Sect.~\ref{iwe}, Sect.~\ref{ime} and Sect.~\ref{inlr}.
\subsection{Integration of the weak eigenstate \label{iwe}}
Starting from \eqn{Lagone} we can construct a manifestly $SU(2)\,\times\,U(1)$ invariant
low energy Lagrangian by integrating out the field $\Pho$ in the limit $\mu_1 \to \infty$. Note 
that, from \eqn{quadmass} the difference between $\mu_1 \to \infty$ and $M_{\mrs} \to \infty$ is
sub-leading in $M_{\mrs}$. This is what has been discussed in 
\Brefs{Gorbahn:2015gxa,Brehmer:2015rna} and we only repeat the observation of 
\Brefs{Gorbahn:2015gxa} that this approach reproduces the effect of scalar mixing on
interactions involving one Higgs scalar $\Ph$, but fails otherwise.
\subsection{Integration of the mass eigenstate \label{ime}}
In the limit $\Lambda = M_{\mrs} \to \infty$ the structure of the calculation is more complex 
since the Lagrangian is given by a power expansion even before integrating out the $\PH$ field,
see \eqn{Lagtwo}. In the following we will describe the steps that are needed to consistently
perform the limit.
\subsubsection{Tadpoles \label{Tad}}
\begin{figure}[t]
   \centering
   \includegraphics[width=0.8\textwidth, trim = 30 250 50 80, clip=true]{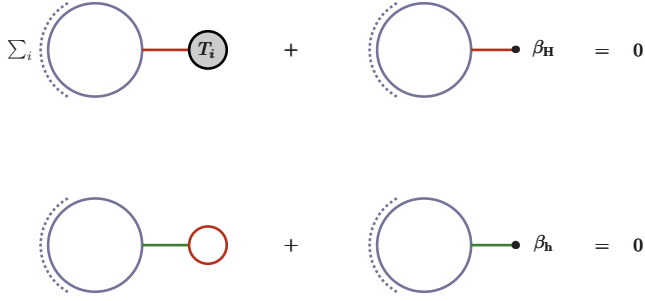}
\vspace{-6.5cm}
\caption[]{Cancellation of tadpoles. Solid (red) lines denote the heavy $\PH$ fields, solid
(green) lines denote the light $\Ph$ field. Balck blobs denote a $\beta\,$-vertex.}
\label{Tadpoles}
\end{figure}

Unless the calculation of observables is performed at tree level, tadpoles should be
introduced and discussed. Their presence and the heavy-light mixing represent an additional 
complication. For instance, in the full singlet extension we have $\PH$ tadpoles and
the relations presented in Sect.~\ref{NandC} must be modified. Thus, working in the
$\beta_{\mathrm{h}}\,$-scheme of \Bref{Actis:2006ra}, we write
\bq
\mu^2_2 = - 2\,\frac{\lambda_2}{g^2}\,M^2 - 2\,\frac{\lambda_{12}}{g^2}\,M^2_{\mrs} +
          \beta_2 \spc
\qquad 
\mu^2_1 = - 2\,\frac{\lambda_1}{g^2}\,M^2_{\mrs} - 2\,\frac{\lambda_{12}}{g^2}\,M^2 +
          \beta_1 \spp
\label{quadmassT}
\eq
Furthermore, we define expansions as follows: 
\bq
\beta_i = \frac{g^2}{16\,\pi^2}\,\beta'_i\,M^2_{\mrs} \spc
\qquad
\beta'_i= \beta^{(0)}_i + \sum_{n=1}\,\beta^{(n)}_i\, \left( \frac{M^2}{M^2_{\mrs}} \right)^n  \spp
\eq
The $\PH$ tadpoles are easily computed in the full theory, giving
\bqa
\mrT_{\PH} &=& - i\,\pi^2\,g\,M\,\sa\,\Bigl[
      2\,M^2 + \frac{M^2}{\ctWq} + 
      \lpar \frac{1}{2}\,\frac{M^2_{\PH}}{M^2} + 3 \rpar\,M^2\,\mrA_0(M) + 
      \lpar \frac{1}{4}\frac{M^2_{\PH}}{M^2} + \frac{3}{2}\,\frac{1}{\ctWs} \rpar\,
      \frac{M^2}{\ctWs}\,\mrA_0(M_0) 
\nl
{}&+& \frac{1}{4}\,\ca\,\lpar 2\,M^2_{\Ph} + M^2_{\PH} \rpar\,
      \lpar \frac{\ca}{M^2} + \frac{\sa}{M\,M_{\mrs}} \rpar \,M^2_{\Ph}\,\mrA_0(M_{\Ph}) + 
      \frac{3}{4}\,\frac{M^2_{\PH}}{\sa}\,
      \lpar \frac{\sa^3}{M^2} + \frac{\ca^3}{M\,M_{\mrs}} \rpar\,M^2_{\PH}\,\mrA_0(M_{\PH}) +
      \mrT^{\Pf}_{\PH}
      \Bigr] \spc
\eqa
where $\mrT^{\Pf}_{\PH}$ is the part induced by fermion loops and $M_0 = M/\ctW$. The constants
$\beta'_i$ are used to cancel $\mrT_{\PH}$. Tadpoles cancellation is illustrated in
Fig.~\ref{Tadpoles}.

The first step in handling tadpoles requires to fix the coefficients $\beta^{(n)}$ so that
$\mrT_{\PH}$ is canceled.
Furthermore, when the $\PH$ field is integrated out we will have to differentiate the $\Ph$ 
tadpoles, those due to a $\PH$ (heavy) loop and those due to loops of light particles; therefore, 
the constants $\beta'_2$ is split into a part that cancels $\PH$ tadpole-loops and a part which 
will be used in performing loop calculations in the low energy theory. We derive
\bqa
\beta^{(0)}_1 &=&
       - 6
         \,t^2_1
         \,\mrA_0(M_{\PH}) \spc
\nl
\beta^{(1)}_1 &=&
       - 2
         \,\lpar t_3 + 2\,t_1 \rpar
         \,\frac{t^2_3}{t_1}
         \,\mrA_0(M_{\PH})
       + \mrT_1 \spc
\nl
\beta^{(2)}_1 &=&
       - \frac{1}{2}
         \,\lpar 2\,t_3 - t_1 \rpar
         \,\frac{t^2_3}{t^2_1}\,x_{\Ph}
         \,\mrA_0(M_{\PH})
       - \frac{1}{4}
         \,\Bigl[ t_1\,x_{\Ph} + 2\,\lpar 2\,t^2_3 + 3\,t_1 \rpar \Bigr]
         \,\frac{1}{t^2_1}
         \,\mrT_1
       + \mrT_2 \spc
\nl
\beta^{(0)}_2 &=&
       - 2 \,\lpar 2\,t_3 + t_1 \rpar
         \,t_3
         \,\mrA_0(M_{\PH}) \spc
\nl
\beta^{(1)}_2 &=&
         \Delta\,\beta^{(1)}_2 
       - \frac{1}{2}
         \,\lpar 12\,t^2_3 + 5\,t_1\,x_{\Ph} \rpar
         \,\frac{t^2_3}{t^2_1}
         \,\mrA_0(M_{\PH})
       + \frac{t_3}{t_1}
         \,\mrT_1 \spc
\nl
\beta^{(2)}_2 &=&
         \Delta\,\beta^{(2)}_2 
       - \frac{1}{2}
         \,\Bigl[ 2\,t_1\,x_{\Ph} + \lpar 7\,t_3 - t_1 \rpar \,t_3 \Bigr]
         \,\frac{t^2_3}{t^3_1}\,x_{\Ph}
         \,\mrA_0(M_{\PH})
       - \frac{1}{2}
         \,\lpar 2\,t^2_3 + 3\,t_1 \rpar
         \,\frac{t_3}{t^3_1}
         \,\mrT_1
       + \frac{t_3}{t_1}
         \,\mrT_2 \spp
\label{Bdef}
\eqa
where the $\mrT$ functions are defined by 
\bqa
\mrT_1 &=&
       - \frac{1}{2}
         \,\mrA_0(M_{\Ph})
         \,x_{\Ph}\,t_3
       - \mrA_0(M)
         \,t_3
       - \frac{1}{2}
         \,\mrA_0(M_0)
         \,\frac{t_3}{\ctWs} \spc
\nl
\mrT_2 &=&
       - \frac{1}{2}
         \,\frac{1}{M^2}\,\mrT^{\Pf}_{\PH}\,\frac{t_3}{t_1}
       - \frac{1}{4}
         \,\mrA_0(M_{\Ph})
         \,\Bigl[ t_1\,x_{\Ph} - \lpar 3\,t_1 - 2\,t_3\,t_1 + 2\,t^2_3 \rpar \Bigr]
         x_{\Ph}\,\frac{t_3}{t^2_1}
\nl
{}&-& \frac{3}{4}
         \,\mrA_0(M_0)
         \,\frac{\stWs}{\ctWq}\,\frac{t_3}{t_1}
       + \frac{1}{4}
         \,\Bigl[ t_1\,x_{\Ph} + 2\,\lpar 3\,t_1 + 2\,t^2_3 \rpar \Bigr]
         \,\frac{1}{t^2_1}\,\mrT_1
       - \frac{1}{2}
         \,\frac{1 + 2\,\ctWq}{\ctWq}
         \,\frac{t_3}{t_1} \spc
\label{Tdef}
\eqa
and where $M^2_{\Ph} = x_{\Ph}\,M^2$.
Working in the $\beta\,$-scheme we have additional loop-induced contributions and
\eqn{Fexp} is modified into
\bqa
\mrF_1(x) &=& \Lambda^3\,\mrFb_1(x) + 
            \Lambda\,\Bigl[ \mrF_{10}(x) + \mrFb_{10}(x) \Bigr] + 
            \Lambda^{-1}\,\Bigl[ \mrF_{11}(x) + \mrFb_{11}(x) \Bigr] + 
            \Lambda^{-3}\,\Bigl[ \mrF_{12}(x) + \mrFb_{12}(x) \Bigr]  \spc
\nl
\mrF_2(x) &=& \Lambda^2\,\mrFb_2(x) + 
            \Bigl[ \mrF_{20}(x) + \mrFb_{20}(x) \Bigr] + 
            \Lambda^{-2}\,\Bigl[ \mrF_{21}(x) + \mrFb_{21}(x) \Bigr] + 
            \Lambda^{-4}\,\Bigl[ \mrF_{22}(x) + \mrFb_{22}(x) \Bigr]  \spc
\label{Fexpm}
\eqa
where the new terms are proportional to $\beta_1$ and $\beta_2$. Therefore, there is an 
additional part in the effective Lagrangian,
\bq
\boxed{
\mcL^{\mrL}_{\eff\,\beta} =
\frac{\Lambda^2}{\xi_0}\,\mcL^{\mrL\,,\,2}_{\eff} +
\frac{1}{\xi^3_0}\,\mcL^{\mrL\,,\,0}_{\eff} +
\frac{1}{\xi^5_0\,\Lambda^2}\,\mcL^{\mrL\,,\,- 2}_{\eff} 
}
\eq
\bqa
\mcL^{\mrL\,,\,2}_{\eff} &=& \mrF_{10}\,\mrFb_1 \spc
\nl\nl
\mcL^{\mrL\,,\,0}_{\eff} &=& 
         \Bigl[
          3\,\mrF_{10}\,\mrF_{30}\,\mrFb_1 + 
         \xi_0\,\mrF_{10}\,\lpar \mrF_{10}\,\mrFb_2 + 2\, \mrF_{20} \,\mrFb_1 - M^2\,\xi_1\,\mrFb_1\rpar + 
         \xi^2_0\,\lpar \mrF_{10}\,\mrFb_{10} + \mrF_{11}\,\mrFb_1 \rpar
           \Bigr] \spc
\nl\nl
\mcL^{\mrL\,,\,- 2}_{\eff} &=& 
     18\,\mrF^3_{10}\,\mrF^2_{30}\,\mrFb_1 + 
     3\,\xi_0\,\mrF^2_{10}\,\mrF_{30}\,\lpar 2\,\mrF_{10}\,\mrFb_2 + 6\,\mrF_{20}\,\mrFb_1 - 
                    3\,M^2\,\xi_1\,\mrFb_1 \rpar 
\nl
{}&+& \xi^2_0\,\Bigl[ \mrF_{10}\,\lpar 3\,\mrF_{10}\,\mrF_{30}\,\mrFb_{10} + 
                3\,\mrF_{10}\,\mrF_{31}\,\mrFb_1 + 
                4\,\mrF_{10}\,\mrF_{20}\,\mrFb_2 - 2\,M^2\,\xi_1\,\mrF_{10}\,\mrFb_2 + 
                6\,\mrF_{11}\,\mrF_{30}\,\mrFb_1 \right.
\nl
{}&+& \left. 4\,\mrF^2_{20}\,\mrFb_1 - 
                4\,M^2\,\xi_1\,\mrF_{20}\,\mrFb_1 \rpar 
          - 2\,\mrFb_1\,\pdmu \mrF_{10}\,\pumu \mrF_{20} \Bigr] 
\nl
{}&+& 
    \xi^3_0\,\Bigl[ \mrF_{10}\,\lpar \mrF_{10}\,\mrFb_{20} + 2\,\mrF_{11}\,\mrFb_2 + 
       2\,\mrF_{20}\,\mrFb_{10} + 2\,\mrF_{21}\,\mrFb_1 - M^2\,\xi_1\,\mrFb_{10} \rpar + 
       2\,\mrF_{11}\,\mrF_{20}\,\mrFb_1 - \pdmu \mrF_{10}\,\pumu \mrFb_{10} \Bigr] 
\nl
{}&+& \xi^4_0\,\lpar \mrF_{10}\,\mrFb_{11} + \mrF_{11}\,\mrFb_{10} \rpar \spc
\eqa
with coefficients $\xi_i$ defined in \eqn{SExidef}.
\subsubsection{Mixed loops \label{Mloop}}
\begin{figure}[t]
   \centering
   \includegraphics[width=0.8\textwidth, trim = 30 250 50 80, clip=true]{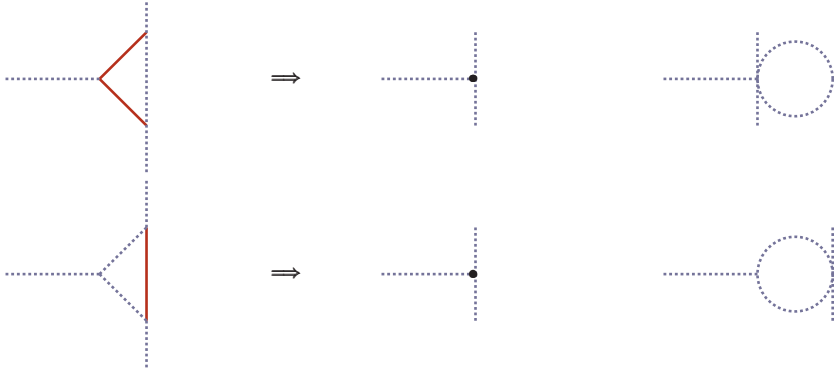}
\vspace{-6.5cm}
\caption[]{Examples of mixed (heavy-light) loops. Solid (red) lines denote heavy fields, dashed
(blue) lines denote light fields. Integrating out the heavy field gives a (contact) local 
operators plus a non-local term which is interpreted as a one-loop diagram in the low energy 
theory.}
\label{MLred}
\end{figure}

In a consistent derivation of the low energy limit we must include also mixed (heavy-light)
loops. Examples of mixed loops are shown in Fig.~\ref{MLred}. Integration of the heavy
fields is performed according to the expansion of three-point functions given in
\appendx{eC0}. Clearly, the result is given by (contact) local operators and by non-local terms
that are one-loop diagrams in the low energy theory, \ie loops with internal light lines.
We give few examples, restricting the external lines to be physical (no $\Ppz, \Pppm$). First 
we define vertices as follows:
\bqa
\mrV^{2}_{\Ph} &=& 2\,g\,M\,t_3\,\lpar 1 - \frac{t_3}{t_1} \rpar \spc
\quad
\mrV^{2}_{\Ph\Ph} = - \frac{1}{2}\,g^2\,t_3 \spc
\quad
\mrV^{10}_{\Ph\Ph} = - 2\,g\,t_3 \spc
\nl
\mrV^{11}_{\Ph\Ph} &=& g\,\frac{t_3}{t_1}\,\lpar 
          M^2\,\frac{t^2_3}{t_1} - 2\,M^2\,t_3 - \frac{3}{2}\,M^2_{\Ph} \rpar \spc
\quad
\mrV^{11}_{\Ph\Ph\Ph} = - \frac{3}{4}\,g^2\,\frac{t_3}{t_1}\,
     \lpar \frac{M^2_{\Ph}}{M} + 4\,M\,\frac{t^2_3}{t_1} - 4\,M\,t_3 \rpar \spc
\nl
\mrV^{11}_{\Ph\PZ\PZ\,;\,\mu\nu} &=& - g^2\,\frac{M}{\ctWs}\,\frac{t_3}{t_1}\,\delta_{\mu\nu} \spc
\quad
\mrV^{11}_{\Ph\PW\PW\,;\,\mu\nu} = - g^2\,M\,\frac{t_3}{t_1}\,\delta_{\mu\nu} \spc
\nl
\mrV^{21}_{\Ph} &=& g\,\frac{t_3}{t^3_1}\,M\,
      \Bigl[ M^2\,t^2_3\,\lpar t_3 - t_1 \rpar + \frac{1}{4}\,M^2_{\Ph}\,
      t_1\,\lpar 3\,t_1 - 5\,t_3 \rpar \Bigr] \spc
\nl
\mrV^{21}_{\Ph\Ph} &=& -\,3\,g^2\,\frac{t^2_3}{t^3_1}\,
      \Bigl[ M^2\,\lpar t_1 - t_3 \rpar^2 + 
      \frac{1}{8}\,M^2_{\Ph}\,t_1 \Bigr] \spc
\nl
\mrV^{21}_{\PZ\PZ\,;\,\mu\nu} &=& - \frac{1}{4}\,g^2\,\frac{M^2}{\ctWs}\,
          \frac{t^2_3}{t^2_1}\,\delta_{\mu\nu} \spc
\quad
\mrV^{21}_{\PW\PW\,;\,\mu\nu} = - \frac{1}{4}\,g^2\,M^2\,\frac{t^2_3}{t^2_1}\,\delta_{\mu\nu} \spc
\quad
\mrV^{30} = - g\,t_1 \spc
\nl
\mrV^{31} &=& g\,M^2\,\frac{t^2_3}{t^2_1}\,\lpar \frac{1}{2}\,t_1 - t_3 \rpar \spc
\quad
\mrV^{31}_{\Ph} = \frac{1}{2}\,g^2\,M\,\frac{t_3}{t_1}\,\lpar t_1 - t_3 \rpar \spc
\nl
\mrV_{\Ph\Ph\Ph} &=&  - \frac{3}{2}\,g\,\frac{M^2_{\Ph}}{M} \spc
\quad
\mrV_{\Ph\Ph\Ph\Ph} = - \frac{3}{2}\,g^2\,\frac{M^2_{\Ph}}{M^2} - 3\,g^2\,\frac{t^2_3}{t_1} \spp
\eqa
As an example we derive the $\Ph^2\,\PZ^2$ (mixed-loop) vertex
\bq
16\,\pi^2\,\mrQ^{\Ph\Ph\PZ\PZ}_{\mu\nu} =
    \frac{1}{8}\,\mrC^{(2)}_0(M_{\Ph})\,
    \lpar \mrV^{2}_{\Ph}\,\mrV^{11}_{\Ph\PZ\PZ} + 
    \mrV^{10}_{\Ph\Ph}\,\mrV^{21}_{\PZ\PZ} \rpar\,\mrV^{10}_{\Ph\Ph}\,\frac{1}{t_1 \Lambda^2}\,
    \delta_{\mu\nu} \spc
\eq
where the scalar three-point function is given in \eqn{Cexpa}.
\subsubsection{Field normalization and parameter shift \label{Cnorm}}
The Lagrangian for the low energy theory requires canonical normalization of the fields which
is a standard procedure when including higher order terms, see
\Brefs{Kallosh:1972ap,Arzt:1993gz,Tyutin:2000ht},
\bq
\upPhi \to \mrZ_{\upPhi}\,\upPhi \spc
\qquad
\mrZ_{\upPhi} = 1 + \frac{g^2}{16\,\pi^2}\,\frac{M^2}{\Lambda^2}\,\Delta \mrZ_{\upPhi} \spp
\eq
In SESM only the $\Ph$ field requires a non-trivial normalization, given by
\bq
\Delta \mrZ_{\Ph} = - \frac{1}{6}\,\frac{t^2_3}{t^3_1}\,\lpar t_1 - t_3 \rpar^2 \spp
\eq
Additionally, we can introduce shifts in the Lagrangian parameters,
\bq
M_{\Ph} = \Bigl[ 1 + \frac{1}{2}\,\frac{g^2}{16\,\pi^2}\,\lpar
\Delta^{(0)}_{M_{\Ph}}\,\frac{\Lambda^2}{M^2} +
\Delta^{(1)}_{M_{\Ph}} +
\Delta^{(2)}_{M_{\Ph}}\,\frac{M^2}{\Lambda^2} \rpar \Bigr]\,{\overline{M}}_{\Ph} \spc
\quad
M = \Bigl[ 1 + \frac{1}{2}\,\frac{g^2}{16\,\pi^2}\,\lpar
\Delta^{(1)}_{M} +
\Delta^{(2)}_{M}\,\frac{M^2}{\Lambda^2} \rpar \Bigr]\,{\overline{M}} \spc
\label{Mshift}
\eq
so that also the bare mass terms (for physical fields) are SM-like. These shifts are given by
\bqa
\Delta^{(0)}_{M_{\Ph}} &=&  - 2\,\frac{t_1 t_3}{x_{\Ph}}\,\mrA_0(M_{\PH}) \spc
\nl 
\Delta^{(1)}_{M_{\Ph}} &=&  - 8\,\,\frac{(t_3 - t_1)^2}{x_{\Ph}}\,t^2_m - 
         \frac{1}{2}
         \,\Bigl[ 3\,t_1\,x_{\Ph} + 4\,\lpar 7\,t^2_3 - 13\,t_1\,t_3 + 7\,t^2_1 \rpar \Bigr]
         \,\frac{t^2_m}{x_{\Ph}}
         \,\mrA_0(M_{\PH}) \spc
\nl
\Delta^{(2)}_{M_{\Ph}} &=& 
        \frac{1}{3}
    \,\Bigl[ 24\,\lpar t_3 - t_1 \rpar\,t^2_3 - \lpar 29\,t_3 - 17\,t_1 \rpar\,t_1\,x_{\Ph} \Bigr]
     \, \lpar t_3 - t_1 \rpar\,\frac{t^2_m}{t^2_1 x_{\Ph}}
\nl
{}&-& \frac{1}{4}
         \,\Bigl[ 3\,t^2_1\,x^2_{\Ph} - 56\,\lpar t_3 - t_1 \rpar^2\,t^2_3 + 2\,
          \lpar 26\,t^2_3 - 43\,t_1\,t_3 + 18\,t^2_1 \rpar \,t_1\,x_{\Ph} \Bigr]
         \,\frac{t^2_m}{t^2_1 x_{\Ph}}
         \,\mrA_0(M_{\PH}) \spc
\nl
\Delta^{(1)}_{M} &=& - \frac{t^2_3}{t_1}\,\mrA_0(M_{\PH}) \spc
\qquad
\Delta^{(2)}_{M} =  - \frac{1}{2}\,\frac{t^2_3}{t^2_1}\,x_{\Ph}\,\mrA_0(M_{\PH}) \spc
\eqa
where we have introduced $t_3= t_m\,t_1$. It is worth noting that the shifted masses
introduced in \eqn{Mshift} remain bare parameters and are not the physical masses.
Furthermore, the shift in $M_{\Ph}$ gives the typical ``fine-tuning'' that is often present when 
we ``derive'' the mass of a low mode (in terms of the scale $\Lambda$) from an UV completion.
\subsection{The complete Lagrangian \label{cL}}
Before introducing the complete Lagrangian we define the concept of (naive) power counting:
any local operator in the Lagrangian is schematically of the form
\bq
\Ope = \frac{M^l}{\Lambda^n}\,{\overline\psi}^a\,\psi^b\,\partial^c\,\lpar \Phi^{\dagger} \rpar^d\,
\Phi^e\,\PA^f \spc
\quad \frac{3}{2}\,(a + b) + c + d + e + f + l - n = 4 \spc
\eq
where Lorentz, flavor and group indices have been suppressed, $\psi$ stands for a generic 
fermion fields, $\Phi$ for a generic scalar and A for a generic gauge field.
All light masses are scaled in units of the (bare) $\PW$ mass $M$. We define dimensions according 
to
\bq
\cdim\,\Ope = \frac{3}{2}\,(a + b) + c + d + e + f \spc
\qquad
\mrdim\,\Ope = \cdim + l \spp
\label{dimdef}
\eq
For a general formulation of power counting see \Bref{Gavela:2016bzc}. The SESM Lagrangian 
can be decomposed as follows,
\bq
\Lag_{\SESM} = \Lag_{\PH=0} + \Lag_{\PH} \spc
\quad
\Lag_{\PH=0} = \Lag_{\mySM}(\Ph) + \sum_{n=0,2}\,\Lambda^{2\,n - 2}\,\Lag_{6 - 2\,n} \spc
\quad
\Lag_{\PH} \to 
\Lag^{\PH}_{\eff} = \Lag^{\mrT}_{\eff} + \Lag^{\mrL}_{\eff} + \Lag^{\beta}_{\eff} \spc
\label{SELag}
\eq
where $\Lag_{\mySM}(\Ph)$ is the SM Lagrangian written in terms of the light Higgs field $\Ph$.
It is worth noting that $\Ph, \PH$ do not transform under irreducible representations of 
$SU(2)\,\times\,U(1)$. 
In \appendx{Lcomplete} we present the full list of operators appearing in $\Lag_{\SESM}$,
classified according to their dimension ($\mrdim= 2,4,6$) and their codimension
($\cdim= 1,\,\dots\,,6$). 
As expected only the SM-like operators acquire coefficients that are $\Lambda\,$-enhanced
($\mrdim = 2$).
The local operators that are usually quoted in this context are $\PKhc$ and
$\pdmuPKhs$ having $\mrdim = \cdim = 6$ ($\PKh = \Ph^2 + \Ppz^2 + 2\,\Ppp\,\Ppm$); however, 
they should not be confused with $\Ope_{\upphi}$ and $\Ope_{\upphi\,\Box}$ of the Warsaw basis 
(see Tab.~2 of \Bref{Grzadkowski:2010es}), the latter being built with a $SU(2)\,\times\,U(1)$ 
scalar doublet while $\PKh$ of \eqn{notI} is not invariant, due to $\Ph$.
\subsubsection{How to use the low energy Lagrangian \label{expla}}
\begin{figure}[t]
   \centering
   \includegraphics[width=0.8\textwidth, trim = 30 250 50 80, clip=true]{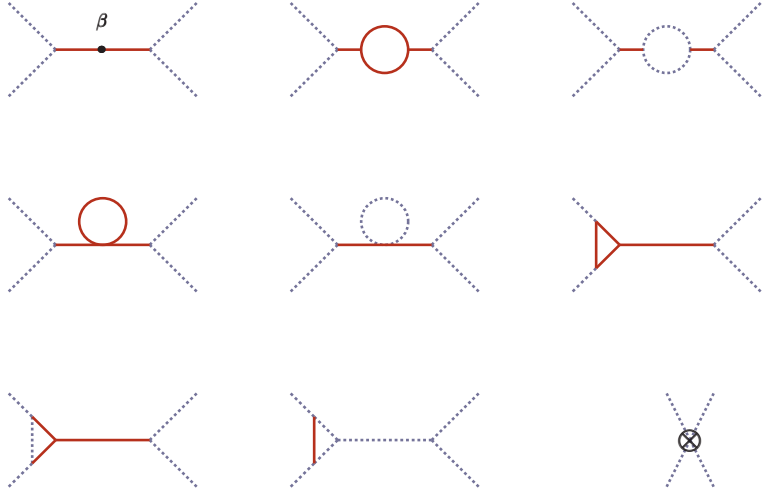}
\vspace{-3.5cm}
\caption[]{Example of a $2 \to 2$ process in the full SESM involving (dashed blue)
light lines. For sake of simplicity we limit the example to scalar lines; moreover,
boxes have not been included and vertex corrections have been shown only for the left part 
of the diagram. Solid (read) lines represent the heavy $\PH$ field and the last diagram 
represents counterterms, both UV and finite (in the ``on-sell'' scheme).}
\label{HSESM}
\end{figure}

The Lagrangian shown in \appendx{Lcomplete} is ready to use but should be used consistently.
No additional problem will arise if we restrict $\Lag_{\SESM}$ to TG operators. When LG
operators are included the following strategy must be adopted. Let us distinguish
between the full theory (HSESM) and the low energy limit (LSESM). In Fig.~\ref{HSESM} 
we show a simple example of a process with four external, light, lines; for the sake of
simplicity we restrict to scalar lines, do not include boxes and avoid the further complication 
due to Dyson resummation. There are loops with (solid red) heavy lines and loops with
(dashed blue) light lines; furthermore, $\beta$ cancels $\PH$ tadpoles, therefore it includes also
light loops. The last diagram in Fig.~\ref{HSESM} includes counterterms, both UV and finite,
UV counterterms are designed to cancel UV poles and by finite counterterms we mean those that
are needed to express bare parameters in terms of experimental quantities (having selected
an input parameter set). Therefore, our scheme is ``on-shell'' (we avoid here complications
induced by using the ``complex-pole'' scheme); the whole procedure is well defined and gauge
parameter independent\footnote{For a discussion on the subtleties induced by the tadpoles
see Sect.~2.4 of \Bref{Actis:2006ra}.}.

\begin{figure}[t]
   \centering
   \includegraphics[width=0.8\textwidth, trim = 30 250 50 80, clip=true]{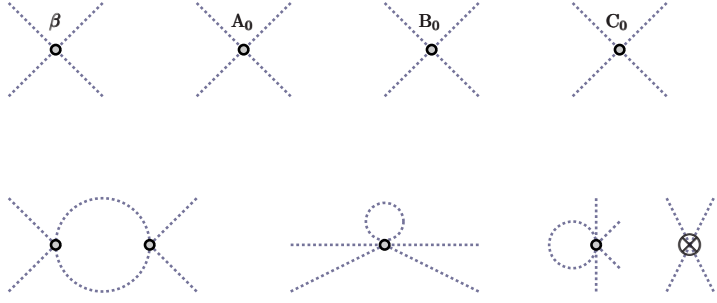}
\vspace{-5.5cm}
\caption[]{The same process as in Fig.~\ref{HSESM} sinn from the low energy side. In the first
row we show diagrams that have been generated by contraction of heavy lines and $\mrA_0, \mrB_0$ 
and $\mrC_0$ keep trace of the origin of the loop contraction, \ie a one-point, two-point and 
three-point loop in HSESM. In the second row we show the diagrams with light loops that have
to be added, including LSESM counterterms in the ``on-shell'' scheme, in order to have
a finite, gauge parameter independent, result.}
\label{LSESM}
\end{figure}

When working in the LSESM framework (at the LG level) the Lagrangian $\Lag_{\SESM}$ will
generate the diagrams in the first row of Fig.~\ref{LSESM}, where dots represent contraction
of $\PH$ propagators. With $\mrA_0, \mrB_0$ and $\mrC_0$ we keep trace of the origin of the
loop contraction, \ie a one-point, two-point and three-point loop in HSESM. To perform a loop
calculation in LSESM we must include light loops, as those shown in the second row of 
Fig.~\ref{LSESM}, taking care of avoiding diagrams that would be two loops in HSESM.
After having included all contributions we take care of renormalization in LSESM by introducing 
UV counterterms and finite counterterms in the ``on-shell'' scheme with a low-energy IPS.
Also this procedure is well defined and gauge parameter independent. Any attempt of performing
a (simpler) $\MSB$ renormalization should be handled with great care. 

To summarize: when the UV completion is known we have a hierarchy among loops in the low-energy
theory. There is a marked contrast between this top-down approach and the bottom-up
effective field theory where one cannot unambiguously identify the powers of hypothetical 
UV couplings present in the Wilson coefficients.
In the EFT approach, by performing the calculations without unnecessary assumptions, it is still 
possible to study the effect of particular hierarchies and specific UV completions (when they 
are precisely defined) a posteriori.
Consider the $h\PZ\PZ$ vertex, we have three contributions (a $\delta_{\mu\nu}$ is left 
understood); 
\bqa
\mrV^{(0)}_{\Ph\PZ\PZ} &=&  - \frac{1}{2}\,\frac{g\,M}{\ctWs} \spc
\qquad
\mrV^{\mrT\mrG}_{\Ph\PZ\PZ} =  \frac{1}{4}\,\frac{g\,M^3}{\Lambda^2\,\ctWs}\,t^2_m \spc
\nl
\mrV^{\mrL\mrG}_{\Ph\PZ\PZ} &=&
        \frac{1}{64}\,\frac{g^3\,M}{\pi^2}\,\frac{t^2_m\,t_1}{\ctWs}\,\mrA_0(M_{\PH})
   + \frac{1}{384}\,\frac{g^3}{\pi^2}\,\frac{M^2}{\Lambda^2\,\ctWs}
         \,\Bigl\{
     3\,\Bigl[ x_{\Ph} - \lpar 18 - 22\,t_m + 7\,t^2_m \rpar\,t_1 \Bigr]\,t^2_m\,\mrA_0(M_{\PH}) 
\nl
{}&+& 3\,\frac{t_m}{t_1}\,\mrT_1 + 
     3\,\frac{t^2_m}{t_1}\,\lpar \beta^{(0)}_2 - \beta^{(0)}_1 \rpar - 
     2\,\lpar 17 - 22\,t_m + 5\,t^2_m \rpar\,t^2_m\,t_1
            \Bigr\}  \spp
\eqa
Here $\mrV^{(0)}_{\Ph\PZ\PZ}$ is SM $\mcO(g)$; $\mrV^{\mrT\mrG}_{\Ph\PZ\PZ}$ is power suppressed, 
$\mcO(g)$ tree-generated; $\mrV^{\mrL\mrG}_{\Ph\PZ\PZ}$ is $\mcO(g^3/\pi^2)$ loop-generated.
Clearly, $\mrV^{\mrT\mrG}_{\Ph\PZ\PZ}$ can be used in any LO/NLO calculation, \ie 
it can be consistently inserted in one loop diagrams containing light particles.
To the contrary, $\mrV^{\mrL\mrG}_{\Ph\PZ\PZ}$ can only be used, at tree level, in one loop
calculations (\ie it should not be inserted into loops of light particles).
Furthermore it is easily seen that mixed-loop contributions to this coupling are 
$\mcO(\Lambda^{-4})$.
\subsection{Gauge invariance \label{GI}}
The Lagrangian under consideration is invariant with respect to the following (infinitesimal)
transformations,
\bqa
\Ph &=&  \Ph + \frac{1}{2}\,g\,\ca\,\lpar \Gamma_{\PZ}\,\frac{1}{\ctW}\,\Ppz + 
         \Ppp\,\Gamma^{\,-} + \Ppm\,\Gamma^{\,+} \rpar \spc
\nl
\PH &=&  \PH + \frac{1}{2}\,g\,\sa\,\lpar \Gamma_{\PZ}\,\frac{1}{\ctW}\,\Ppz + 
         \Ppp\,\Gamma^{\,-} + \Ppm\,\Gamma^{\,+} \rpar \spc
\nl
\Ppz &=&  \Ppz - \frac{1}{2}\,g\,\Gamma_{\PZ}\,\frac{1}{\ctW}\,\lpar \ca\,\Ph + \sa\,\PH + 
          2\,\frac{M}{g} \rpar + 
          \frac{i}{2}\,g\,\lpar \Gamma^{\,-}\,\Ppp - \Gamma^{\,+}\,\Ppm \rpar \spc
\nl
\Ppm &=&  \Ppm - \frac{1}{2}\,g\,\Gamma^{\,-}\,\lpar \ca\,\Ph + \sa\,\PH + 
          2\,\frac{M}{g} + i\,\Ppz \rpar + 
          \frac{i}{2}\,g\,\Bigl[ \lpar \ctW^2 - \stW^2 \rpar\,\frac{1}{\ctW}\,\Gamma_{\PZ} + 
          2\,\stW\,\Gamma_{\PA} \Bigr]\,\Ppm \spc
\nl
\PA_{\mu} &=&  \PA_{\mu} + i\,g\,\stW\,\lpar \Gamma^{\,-}\,\PWp_{\mu} - 
               \Gamma^{\,+}\,\PWm_{\mu} \rpar - \pdmu\,\Gamma_{\PA} \spc
\nl
\PZ_{\mu} &=&  \PZ_{\mu} + i\,g\,\ctW\,\lpar \Gamma^{\,-}\,\PWp_{\mu} - 
               \Gamma^{\,+}\,\PWm_{\mu} \rpar - \pdmu\,\Gamma_{\PZ} \spc
\nl
\PWm_{\mu} &=&  \PWm_{\mu} - i\,g\,\Gamma^{\,-}\,\lpar \ctW\,\PZ_{\mu} + \stW\,\PA_{\mu} \rpar +
                i\,g\,\lpar \ctW\,\Gamma_{\PZ} + \stW\,\Gamma_{\PA} \rpar\,\PWm_{\mu} - 
                \pdmu\,\Gamma^{\,-} \spc
\label{Gtran}
\eqa
when we expand $\sa, \ca$ to any given order. The gauge invariance of the low energy theory
must be understood as follows: the transformations of \eqn{Gtran} may be seen as generating new
vertices in the theory and gauge invariance requires that, for any Green's function, the
sum of all diagrams containing one $\Gamma\,$-vertex cancel.
When sources are added to the Lagrangian the field transformation generates special vertices
that are used to prove equivalence of gauges and simply-contracted Ward-Slavnov-Taylor
identities~\cite{'tHooft:1972ue}.
Therefore, for any ``transformed'' Green's function we integrate the $\PH$ field and,
order-by-order in $\Lambda$, terms containing one $\Gamma\,$-vertex continue to cancel
(and WST identities to be valid). For instance, if we set $\ca = 1$ and $\sa = 0$ in \eqn{Gtran},
it is easily seen that $\Lag_{\PH=0}$, given in \eqn{SELag}, is not invariant but the addition
of $\Lag^{\mrT}_{\eff}$ truncated at $\mcO(1)$ restores gauge invariance.

\subsection{Integration in the non-linear representation \label{inlr}}
An interesting alternative, see \Brefs{Alonso:2015fsp,Gavela:2014uta,Buchalla:2015qju} is 
represented by the following Lagrangian
\bqa
\mcL &=&
 - \frac{1}{2}\,\pdmu \Pht\,\pdmu \Pht       
 - \frac{1}{2}\,F^2(\Pht)\,g_{ab}(\upphi)\,\mrD_{\mu} \upphi^a\,\mrD_{\mu} \upphi^b
 - \pdmu\,\upchi\,\pdmu\,\upchi
 - \frac{1}{2}\,\mu^2_2\,\lpar \Pht + \srt\,\mrv \rpar^2
\nl
{}&-& \mu^2_1\,\upchi^2
 - \frac{1}{8}\,\lambda_2\,\lpar \Pht + \srt\,\mrv \rpar^4
 - \frac{1}{2}\,\lambda_1\,\upchi^4
 - \frac{1}{2}\,\lambda_{12}\,\upchi^2\,\lpar \Pht + \srt\,\mrv \rpar^2 \spc
\label{LagNL}
\eqa
where we have introduced a complex scalar doublet
\bq
\Upphi = \frac{1}{\srt}\,\left(
\begin{array}{c}
\phi^4_{\PH} + i\,\phi^3_{\PH} \\
\phi^2_{\PH} + i\,\phi^1_{\PH} 
\end{array}
\right)
\eq
with ``polar'' coordinates defined by
\bq
\phi^i_{\PH} = \lpar \Pht + \mrv \rpar\,u^i(\phi) \spc
\quad
u(\phi)\,\cdot\,u(\phi) = 1 \spc
\quad
u^i(0) = \delta^{i4} \spp
\eq
where $i,j,\dots = 1,\,\dots\,,4$ and $a,b,\dots = 1,\,\dots\,,3$. A choice for the metric,
present in \eqn{LagNL}, is
\bq
g_{ab}(\phi) = \delta_{ab} + \frac{\phi_a\,\phi_b}{\mrv^2 - \phi_a\,\phi^a} \spc
\eq
and the covariant derivative in \eqn{LagNL} is defined in Eqs.(15-16) of \Bref{Alonso:2015fsp}.
When discussing the SM one uses
\bq
F_{\mySM}\lpar \Pht \rpar = 1 + \frac{\Pht}{\mrv}
\eq
in \eqn{LagNL}. After mixing with the singlet we obtain
\bqa
F^2 &=& F^2_{\mySM}\lpar \Ph \rpar + \frac{2}{\mrv}\,F_{\mySM}\lpar \Ph \rpar\, 
\Bigl[ \lpar \ca - 1 \rpar\,\Ph + \sa\,\PH \Bigr] + \;\dots
\nl
{}&=& \Bigl\{ 1 + \frac{\Ph}{\mrv}\,
      \Bigl[ 1 - \frac{1}{2}\,\frac{t^2_3}{t^2_1}\,\frac{M^2}{M^2_{\mrs}} +
      \mcO \lpar M^{-4}_{\mrs} \rpar \Bigr] + 
      \frac{\PH}{\mrv}\,\frac{t_3}{t_1}\,\frac{M}{M_{\mrs}} 
      + \mcO \lpar \PH^2 \rpar
      \Bigr\}^2 
\nl
{}&=& \Bigl[ 1 + f_{\Ph}\,\frac{\Ph}{\mrv} + f_{\PH}\,\frac{\PH}{\mrv} + 
             \mcO \lpar \PH^2 \rpar \Bigr]^2\spp
\eqa
After integrating $\PH$ we have two effects, a change in \eqn{LagNL} due to a redefinition
of $F$ and the addition of higher dimensional operators, \eg
\bq
f^2_{\PH}\,\lpar 1 + f_{\Ph}\,\frac{\Ph}{\mrv} \rpar^2\,\mcO^2_{\phi} \spc
\qquad
\mcO_{\phi} = g_{ab}(\upphi)\,\mrD_{\mu} \upphi^a\,\mrD_{\mu} \upphi^b \spp
\eq
Note that $f_{\Ph} \not = 1$ has an effect on curvatures, see Eq.(22) and Eq.(27) of
\Bref{Alonso:2015fsp}. From this point of view the geometric formulation of the Higgs EFT seems
the most promising road to account for general mixings in the scalar sector.
\section{Low energy behavior for THDM models \label{EFTD}}
We consider THDM with softly-broken $\mrZ_2$ 
symmetry~\cite{Branco:2011iw,Yagyu:2012qp,Bhattacharyya:2015nca}. The bosonic part of the 
Lagrangian is given by
\bqa
\mcL &=& - \sum_{i=1,2}\,\lpar \mrD_{\mu}\,\Upphi_i \rpar^{\dagger}\,\mrD_{\mu}\,\Upphi_i
      + \sum_{i=1,2}\,\mu^2_i\,\Upphi^{\dagger}_i\,\Upphi_i
      + \mu^2_3\,\lpar \Upphi^{\dagger}_1\,\Upphi_2 + \Upphi^{\dagger}_2\,\Upphi_1 \rpar
\nl
{}&+& \frac{1}{2}\,\sum_{i=1,2}\,\lambda_i\,\lpar \Upphi^{\dagger}_i\,\Upphi_i \rpar^2
      + \lambda_3\,\Upphi^{\dagger}_1\,\Upphi_1\,\Upphi^{\dagger}_2\,\Upphi_2
      + \lambda_4\,\Upphi^{\dagger}_1\,\Upphi_2\,\Upphi^{\dagger}_2\,\Upphi_1
      + \frac{1}{2}\,\lambda_5\,\Bigl[
        \lpar \Upphi^{\dagger}_1\,\Upphi_2 \rpar^2 +
        \lpar \Upphi^{\dagger}_2\,\Upphi_1 \rpar^2 \Bigr] \spp
\eqa
With doublets given by
\bq
\Upphi_i = \frac{1}{\srt}\,\left(
\begin{array}{c}
\Ph_i + \srt\,\mrv_i + i\,\Ppz \\
\srt\,i\,\Ppm
\end{array}
\right)
\eq
The mixing angle $\beta$ is such that
\bqa
\Ph_1 &=& - \mrv_1 + \cbe\,\lpar \Ph'_1 + \mrv \rpar  - \sbe\,\Ph'_2 \spc
\quad
\upphi^0_1 = \cbe\,\upphi^0 - \sbe\,\PA^0 \spc
\quad
\upphi^{\pm}_1 = \cbe\,\upphi^{\pm} - \sbe\,\PH^{\pm}
\nl
\Ph_2 &=& - \mrv_2 + \sbe\,\lpar \Ph'_1 + \mrv \rpar  - \cbe\,\Ph'_2 \spc
\quad
\upphi^0_2 = \sbe\,\upphi^0 + \sbe\,\PA^0 \spc
\quad
\upphi^{\pm}_2 = \sbe\,\upphi^{\pm} + \cbe\,\PH^{\pm} \spp
\eqa
with $\mrv^2 = \mrv^2_1 + \mrv^2_2$. Finally, diagonalization in the neutral sector gives
\bq
\Ph'_1 =  \cab\,\PH - \sab\,\Ph \spc
\qquad
\Ph'_2 =  \sab\,\PH + \cab\,\Ph \spp
\eq
The first problem in deriving the low energy behavior is represented by the individuation
of the cutoff scale. In the unbroken phase one can use the Plehn scale
\bq
\Lambda^2 = \mu^2_1\,\sbes + \mu^2_2\,\cbes + \mu^2_3\,\sbe \spc
\eq
whereas in the mass eigenstates, \Bref{Brehmer:2015rna} suggests $M^2_{\PA}$, based on the
fact that custodial symmetry requires almost degenerate heavy states. Our procedure
is as follows. First we eliminate $\mu^2_{1,2}$ by means of the following transformation:

\bqa
\cbes\,\mu^2_1 &=&
          \beta_1
          - \mrv^2\,\Bigl[ \sbes\,\cbes\,\lb + \frac{1}{2}\,
                          \lpar \lambda_2\,\sbeq + \lambda_1\,\cbeq \rpar \Bigr]
          - 2\,\sbe\,\cbe\,\mu^2_3
          +  \sbes\,\mu^2_2 \spc
\nl
\cot\beta\,\mu^2_2 &=&
          \beta_2
          - \tbe\,\beta_1
          + \frac{1}{2}\,\mrv^2\,\Bigl[
           \sbe\,\cbe\,\lb +  \lpar \tbe - \sbe\cbe \rpar \Bigr]
          + \mu^2_3 \spc
\eqa
where $\beta_{1,2}$ are the constants needed to cancel tadpoles and
$\lb = \lambda_3 + \lambda_4 + \lambda_5$. Next we write
\bq
\mu^2_3 = \sbe\,\cbe\,\Msbs \spc
\quad
\mrv^2\,\lambda_5 = M^2_{\PAz} - \Msbs \spc
\quad
\mrv^2\,\lambda_4 = 2\,\beta_1 + 2\,\frac{\cbes - \sbes}{\sbe \cbe}\,\beta_2 + 
                    2\,M^2_{\PHpm} - M^2_{\PAz} - \Msbs \spp
\eq
\bqa
\mrv^2\,\lambda_2 &=& \mrv^2\,\lpar 2\,\lb - \lambda_1 \rpar + 
                   {\Msbs - M^2_{22}}{\sbes \cbes} \spc
\nl
\mrv^2\,\lb &=& \frac{1}{2}\,\lpar \mrv^2\,\lambda_1\,\frac{\sbeq - \cbeq}{\sbes} 
          + \frac{M^2_{22} - \Msbs}{\cbes}
          - \frac{M^2_{11}}{\sbes} \rpar \spc
\nl
\mrv^2\,\lambda_1 &=& 2\,\tbe\,M^2_{12} + 
                      \tbes\,\lpar \Msbs - M^2_{22} \rpar - M^2_{11} \spc
\eqa
Requiring
\bq
\sab\,\cab\,\lpar M^2_{11} - M^2_{22} \rpar + \Bigl[\sabs - \cabs\Bigr] = 0 \spc
\eq
gives the following result for the neutral masses:
\bq
M^2_{\Ph} = M^2_{11} - \cot(\alpha - \beta)\,M^2_{12} \spc
\qquad
M^2_{\PH} = M^2_{12} + \sab\,\cab\,M^2_{\Ph}
\eq
Our scenario or the THDM is defined by $\Lambda = \Msb  >> \mrv$ and
\bq
\beta = \frac{1}{2}\,\lpar \pi - \dbe \rpar \spc
\qquad
\alpha = \frac{1}{2}\,\dal \spp
\eq
Expanding in $\mrv/\Lambda$ we obtain:
\bq
\lambda_2 = - \frac{M^2_{\Ph}}{\mrv^2} - 
            \frac{1}{2}\,\lpar M^2_{\Ph} + \mrv^2\,\lb \rpar\,\frac{\mrv^2}{\Lambda^4} \spc
\quad
\dbe =  \frac{\mrv^2}{\Lambda^2} + \mcO\lpar \frac{\mrv^4}{\Lambda^4} \rpar \spc
\quad
\dal =  - \frac{\mrv^2}{\Lambda^2} + \mcO\lpar \frac{\mrv^4}{\Lambda^4} \rpar \spp
\eq
All masses and angles are re-expressed in term of $\Lambda$ and couplings.
\bqa
M^2_{\PHpm} &=& \Lambda^2 + \frac{1}{2}\,\mrv^2\,( \lambda_4 + \lambda_5 ) \spc
\quad
M^2_{\PAz} = \Lambda^2 + \mrv^2\,\lambda_5 \spc
\quad
M^2_{\PH} = \Lambda^2 - \frac{1}{4}\,\Bigl[
\mrv^2\,\lpar \lambda_1 - 2\,\lb \rpar - M^2_{\Ph} \Bigr]\,\frac{\mrv^4}{\Lambda^4}
\nl
\sbe &=& 1 - \frac{1}{8}\,\frac{\mrv^4}{\Lambda^4}
       + \mcO\lpar \frac{\mrv^6}{\Lambda^6} \rpar \spc 
\qquad
\cbe = \frac{1}{2}\,\frac{\mrv^2}{\Lambda^2}
             + \mcO\lpar \frac{\mrv^4}{\Lambda^4} \rpar \spc
\nl
\sab &=& - 1 + \mcO\lpar \frac{\mrv^6}{\Lambda^6} \rpar \spc
\qquad
\cab = -\,\frac{1}{2}\,\lpar M^2_{\Ph} + \mrv^2\,\lb \rpar \frac{\mrv^2}{\Lambda^4}
             + \mcO\lpar \frac{\mrv^6}{\Lambda^6} \rpar \spp
\label{THDMexp}
\eqa
Using \eqn{THDMexp} we can expand the Lagrangian in powers of $\Lambda^{-1}$ and apply
the formalism of Sect.~\ref{Gform} to obtain the low energy limit of the model.
Also for THDM models the SM decoupling limit cannot be obtained by making only assumptions 
about one parameter. For a general discussion on alignment and decoupling, 
see~\Brefs{Gunion:2002zf,Carena:2013ooa}.

There are four THDM models that differ in the fermion sector: they are type I, II, X and Y,
see \Bref{Yagyu:2012qp} for details. The THDM Lagrangian becomes
\bq
\Lag_{\THDM} = \Lag_{\THDM}\bmid_{\mbox{heavy}=0} + \Lag^{\mbox{heavy}}_{\THDM} \spc
\quad
\Lag_{\THDM}\bmid_{\mbox{heavy}=0} = \Lag_0 + \Lambda^{-2}\,\Lag_2 \spc
\eq
with $\Lag_2 = 0$ for THDM type I. The heavy part of the Lagrangian,
$\Lag^{\mbox{heavy}}_{\THDM}$ is given by a sum of terms; we define the set
$\{\upPhi\} = \{\PH,\PAz,\PHpm\}$ and obtain
\bq
\Lag^{\mbox{heavy}}_{\THDM} = \sum_n\,\Lag^{\mbox{heavy}}_{n} \spc
\eq
with
\bq
\Lag^{\mbox{heavy}}_1 = \sum_{\upphi \in \{\upPhi\}}\,\mrF_{1\,\upphi}\,\upphi \spc
\qquad
\mrF_{1\,\PHm} = \lpar \mrF_{1\,\PHp} \rpar^{\dagger} \spc
\label{Lterm}
\eq
\bq
\Lag^{\mbox{heavy}}_2 = \sum_{\upphi_i,\upphi_j \in \{\upPhi\}}\,
                        \upphi_i\,\mrF_{2\,\upphi_i\upphi_j}\,\upphi_j \spc
\label{Qterm}
\eq
where $\mrF_{2\,\upphi_i\upphi_j}$ contains derivatives and where terms with one or two
heavy fields are of $\mcO(1)$ and $\mcO(\Lambda^{-2})$. With three fields we have
\bq
\Lag^{\mbox{heavy}}_3 = \sum_{\upphi_i,\upphi_j,\upphi_k \in \{\upPhi\}}\,
                        \mrF_{3\,\upphi_i\upphi_j\upphi_k}\,\upphi_i\upphi_j\upphi_k \spc
\eq
where $\mrF_3$ is $\mcO(\Lambda^{-2})$. Finally we have
\bq
\Lag^{\mbox{heavy}}_4 = \sum_{\upphi_i,\upphi_j,\upphi_k,\upphi_l \in \{\upPhi\}}\,
                        \mrF_{4\,\upphi_i\upphi_j\upphi_k\upphi_l}\,
                        \upphi_i\upphi_j\upphi_k\upphi_l \spc
\eq
with contributions of $\mcO(1)$ and $\mcO(\Lambda^{-2})$. For THDM type II, X and Y there
are terms of $\mcO(\Lambda^2)$ in $\mrF_1$.

Due to the SM-like scenario, $\sab = - 1 + \mcO(\mrv^6/\Lambda^6)$, $\Ph$ is almost the SM 
Higgs boson (alignment). If we consider the vertex $\Ph\PGg\PGg$ the only
deviation (at $\mcO(\Lambda^{-2})$) is given by the $\PHpm$ loops which, after expansion,
contribute to the ``contact'' term
\bq
\delta \mrV_{\Ph\PGg\PGg} = \frac{1}{3}\,i\,g^3\,\lpar 2\,\lambda_5 - \lambda_3 \rpar\,\stW\,
\frac{M}{\Lambda^2}\,\lpar 2\,p^{\mu}_2\,p^{\nu}_1 + M^2_{\Ph}\,\delta^{\mu\nu} \rpar \spc
\eq
and there is no contribution from insertion of local operators into SM loops.

There are constraints from electroweak precision data, most noticeably from the $\rho$ parameter. 
The contribution from scalar loops in THDM is
\bqa
\Delta \rho_{\THDM} &=& \frac{\myGF}{8\,\sqrt{2}\,\pi^2}\,\Bigl\{
\mrF\lpar M_{\PAz}\,,\,M_{\PHpm} \rpar
- \cabs\,\Bigl[ \mrF\lpar M_{\Ph}\,,\,M_{\PAz} \rpar  - 
                \mrF\lpar M_{\Ph}\,,\,M_{\PHpm} \rpar \Bigr]
\nl
{}&-& \sabs\,\Bigl[ \mrF\lpar M_{\PH}\,,\,M_{\PAz} \rpar  - 
                \mrF\lpar M_{\PH}\,,\,M_{\PHpm} \rpar \Bigr] \Bigr\} \spc
\nl
\mrF\lpar m_1\,,m_2 \rpar &=& \frac{1}{2}\,\lpar m^2_1 + m^2_2 \rpar -
\frac{m^2_1\,m^2_2}{m^2_1 - m^2_2}\,\ln\frac{m^2_1}{m^2_2} \spp
\eqa
In the scenario described by \eqn{THDMexp} we obtain 
\bq
\Delta \rho_{\THDM} = \frac{\myGF}{96\,\sqrt{2}\,\pi^2}\,\frac{\mrv^4}{\Msbs}\,
\lpar \lambda^2_4 - \lambda^2_5 \rpar \spc
\eq
\ie a mass suppressed correction. Deriving TG operators is relatively easy; using \eqn{Lterm}
and neglecting quadratic terms, \eqn{Qterm}, we define
\bq
\mrF_{1\,\upphi} = \mrF^0_{1\,\upphi} + \Msb^{\,-2}\,\mrF^2_{1\,\upphi} \spc
\eq
and derive the following result
\bq
\Lag^{\mrT\mrG}_{\THDM} = \frac{1}{2}\,\Msb^{\,-2}\,\Lag^{\mrT\mrG}_{\THDM\,2} + 
                                       \Msb^{\,-4}\,\Lag^{\mrT\mrG}_{\THDM\,4} \spc
\eq
\bqa
\Lag^{\mrT\mrG}_{\THDM\,2} &=& 
       \lpar \mrF^0_{1\,\PH} \rpar^2 +
       \lpar \mrF^0_{1\,\PAz} \rpar^2 +
       2\,\mrF^0_{1\,\PHp}\,\mrF^0_{1\,\PHm} \spc
\nl\nl
\Lag^{\mrT\mrG}_{\THDM\,4} &=& 
          \mrF^0_{1\,\PH}\,\mrF^2_{1\,\PH} +
          \mrF^0_{1\,\PAz}\,\mrF^2_{1\,\PAz} +
          \mrF^0_{1\,\PHp}\,\mrF^2_{1\,\PHm} +
          \mrF^0_{1\,\PHm}\,\mrF^2_{1\,\PHp} 
\nl
{}&-&
    \frac{1}{2}\,\pdmu \mrF^0_{1\,\PH}\,\pdmu \mrF^0_{1\,\PH} -
    \frac{1}{2}\,\pdmu \mrF^0_{1\,\PAz}\,\pdmu \mrF^0_{1\,\PAz} -
                 \pdmu \mrF^0_{1\,\PHp}\,\pdmu \mrF^0_{1\,\PHm} 
\nl
{}&-& \,\frac{1}{2}\,g^2\,\mrv^2\,\Bigl[
        ( t_4 + t_5 )\,\mrF^0_{1\,\PHp} + \mrF^0_{1\,\PHm} +
          t_5\,\lpar \mrF^0_{1\,\PAz} \rpar^2 \Bigr] \spp
\eqa
Note that $\mrF^0_{1\,\upphi} = 0$ if we do not include $\beta$ terms; we derive
($\PKz = \Ppz^2 + 2\,\Ppp\,\Ppm$)
\bqa
\mrF_{1\,\PH} &=&
       - \frac{g\,M}{\Lambda^2} \, \Bigl[
          \frac{1}{2}\,M^2\,\lpar x_{\Ph} + 4\,\tb \rpar\,\lpar 3\,\Ph^2 + \PKz \rpar
          + M_{\Pl}\,\PAl\,\Pl + M_{\PQu}\,\PAQu\,\PQu + M_{\PQd}\,\PAQd\,\PQd
          \Bigr]
\nl
{}&-& \frac{1}{4}\,\frac{g^2\,M^2}{\Lambda^2} \, 
          \lpar x_{\Ph} + 4\,\tb \rpar\,\Ph\,\lpar \Ph^2 + \PKz \rpar \spc
\nl\nl
\mrF_{1\,\PAz} &=&
        \frac{g\,M^3}{\Lambda^2} \, 
          \lpar x_{\Ph} + 4\,\tb \rpar\,\Ph\,\Ppz
       + \frac{1}{4}\,\frac{g^2\,M^2}{\Lambda^2} \, 
          \lpar x_{\Ph} + 4\,\tb \rpar\,\Ppz\,\PKh
\nl
{}&+& i\,\frac{g\,M}{\Lambda^2} \, \Bigl[
        M_{\Pl}\,\PAl\,\gamma^5\,\Pl - M_{\PQu}\,\PAQu\,\gamma^5\,\PQu + 
        M_{\PQd}\,\PAQd\,\gamma^5\,\PQd
          \Bigr] \spc
\nl\nl
\mrF_{1\,\PHm} &=&
        \frac{g\,M^3}{\Lambda^2} \, 
         \lpar x_{\Ph} + 4\,\tb \rpar\,\Ph\,\Ppp       
       + \frac{1}{4}\,\frac{g^2\,M^2}{\Lambda^2} \, 
          \lpar x_{\Ph} + 4\,\tb \rpar\,\Ppp\,\PKh
\nl
{}&-& i\,\frac{g\,M}{\srt\,\Lambda^2} \, \Bigl[
        M_{\Pl}\,\PAl\,\gamma_{+}\,\PGnl + M_{\PQd}\,\PAQd\,\gamma_{+}\,\PQu - 
        M_{\PQu}\,\PAQd\,\gamma_{-}\,\PQu
          \Bigr] \spc
\nl\nl
\mrF_{1\,\PHp} &=&
        \frac{g\,M^3}{\Lambda^2} \, 
          \lpar x_{\Ph} + 4\,\tb \rpar\,\Ph\,\Ppm
       + \frac{1}{4}\,\frac{g^2\,M^2}{\Lambda^2} \, 
          \lpar x_{\Ph} + 4\,\tb \rpar\,\Ppm\,\PKh
\nl
{}&+& i\,\frac{g\,M}{\srt\,\Lambda^2} \, \Bigl[
        M_{\Pl}\,\PAGnl\,\gamma_{-}\,\Pl - M_{\PQu}\,\PAQu\,\gamma_{+}\,\PQd + 
        M_{\PQd}\,\PAQu\,\gamma_{-}\,\PQd
          \Bigr] \spc
\eqa
where $\Lambda^2= g^2\,\Msbs$.
Finally, in the limit described by \eqn{THDMexp} LG operator are more abundant than TG ones,
one-point functions are $\mcO(\Msb^2)$, two-point functions are $\mcO(1)$ and 
three-point functions are $\mcO(\Msb^{\,-2})$. They all involve internal (loop) heavy lines while 
at tree level any heavy line is quadratically suppressed.
To give an example we split the $\mrF_2$ functions as follows:
\bq
\mrF_{2\,ij} = \mrF^0_{2\,ij} + \Lambda^{-2}\,\mrF^2_{2\,ij} + \mcO(\Lambda^{-4}) \spc
\qquad
\mrF^0_{2\,ij} = \mrF^{00}_{2\,ij} + 
\lpar \mrF^{0\mu}_{2\,ij}\,\drmu - \dlmu\,\mrF^{0\mu}_{2\,ij} \rpar \spc
\eq
\bqa
\mrF_{2\,\PH\PH} &=& g\,M\,\tb\,\Ph
       + \frac{1}{8}\,g^2 \, \Bigl[
           2\,\tb\,\PKh
          - 4\,t_5\,\Ppz^2
          - 4\,( t_4 + t_5 )\,\Ppp\,\Ppm
          - \PZ^2\,\frac{1}{\ctWs}
          - 2\,\PWp^{\mu}\,\PWm_{\mu}
          \Bigr]
\nl
{}&+& \frac{1}{8}\,g^3\,\frac{M^3}{\Lambda^2}\,\mrT_c\,\Ph
       + \frac{1}{32}\,g^4\,\frac{M^2}{\Lambda^2}\,\mrT_c\,\PKh \spc
\nl\nl
\mrF_{2\,\PAz\PAz} &=& - g\,M \, (2\,t_5 - \tb)\,\Ph
       + \frac{1}{8}\,g^2 \, \Bigl[
           2\,\tb\,\PKh
          - 4\,t_5\,\Ph^2
          - 4\,( t_4 + t_5 )\,\Ppp\,\Ppm
          - \PZ^2\,\frac{1}{\ctWs}
          - 2\,\PWp^{\mu}\,\PWm_{\mu}
          \Bigr]
\nl
{}&+& \frac{1}{8}\,g^3\,\frac{M^3}{\Lambda^2}\,\mrT_c\,\Ph
       + \frac{1}{32}\,g^4\,\frac{M^2}{\Lambda^2}\,\mrT_c\,\PKh \spc
\nl\nl
\mrF_{2\,\PH\PAz} &=& - 2\,g\,M\,t_5\,\Ppz
          - g^2\,t_5\,\Ppz\,\Ph
          + \frac{1}{2}\,g\,(\PZ^{\mu}\,\drmu - \dlmu\,\PZ^{\mu})\,\frac{1}{\ctW} \spc
\nl\nl
\mrF_{2\,\PH\PHm} &=& - g\,M\,( t_4 + t_5 )\,\Ppp
          - \frac{1}{2}\,g^2\,( t_4 + t_5 )\,\Ppp\,\Ph
          + \frac{1}{2}\,i\,g^2 \, \Bigl[
           ( t_5 - t_4 )\,\Ppz\,\Ppp
          + \PZ^{\mu}\,\PWp_{\mu}\,\frac{\stWs}{\ctW}
          - \PA^{\mu}\,\PWp_{\mu}\,\stW
          \Bigr]
\nl
{}&+& \frac{1}{2}\,g\,(\PWp^{\mu}\,\drmu - \dlmu\,\PWp^{\mu}) \spc
\nl\nl
\mrF_{2\,\PAz\PHm} &=&
        \frac{1}{2}\,g^2 \, \Bigl[
           ( t_4 + t_5 )\,\Ppz\,\Ppp
          - \PZ^{\mu}\,\PWp_{\mu}\,\frac{\stWs}{\ctW}
          + \PA^{\mu}\,\PWp_{\mu}\,\stW
          \Bigr]
       + i\,g\,M\,( t_5 - t_4 )\,\Ppp
\nl
{}&+& \frac{1}{2}\,i\,g^2\,( t_5 - t_4 )\,\Ppp\,\Ph
       + \frac{1}{2}\,i\,g\,(\PWp^{\mu}\,\drmu - \dlmu\,\PWp^{\mu}) \spc
\nl\nl
\mrF_{2\,\PHm\PHm} &=& \frac{1}{2}\,g^2\,t_5\,\Ppp\,\Ppp \spc
\nl \nl
\mrF_{2\,\PHp\PHm} &=&
        2\,g\,M \, \Bigl[
           \tb
          - 2( t_4 + t_5 )
          \Bigr]\,\Ph
\nl
{}&+& \frac{1}{4}\,g^2 \, \Bigl[
           2\,\tb\,\PKh
          - 2\,( t_4 + t_5 )\,\Ph^2
          - 2\,( t_4 + t_5 )\,\Ppz^2
          - 4\,t_5\,\Ppp\,\Ppm
          - \frac{1 - 4\,\stWs\,\ctWs}{\ctWs}\,\PZ^2
\nl
{}&+& 4\,(1 - 2\,\ctWs)\,\frac{\stW}{\ctW}\,\PZ^{\mu}\,\PA_{\mu}
          - 4\,\stWs\,\PA^2
          - \PWp^{\mu}\,\PWm_{\mu}
          \Bigr]
\nl
{}&+& \frac{1}{4}\,g^3\,\frac{M^3}{\Lambda^2}\,\mrT_c\,\Ph
       + \frac{1}{16}\,g^4\,\frac{M^2}{\Lambda^2}\,\mrT_c\,\PKh
       + \frac{1}{2}\,i\,g \, 
           \frac{1 - 2\,\ctWs}{\ctW}\,
          (\PZ^{\mu}\,\drmu - \dlmu\,\PZ^{\mu})
       - i\,g\,\stW\,
          (\PA^{\mu}\,\drmu - \dlmu\,\PA^{\mu}) \spp
\eqa
where we have introduced
\bq
\mrT_c =  128\,( t_1 - \tb )^2\,\frac{t_1}{x_{\Ph} + 4\,t_1}
          - 3\,x^2_{\Ph}
          - 32\,( t_1 - \tb )^2
          - 4\,(5\,\tb - 2\,t_1)\,x_{\Ph} \spp
\eq
To give an example we show the loop operators generated by integrating heavy bubbles:
\bq
\Lag_b = \frac{1}{16\,\pi^2}\,\lpar \Lag^0_b + \frac{1}{\Lambda^2}\,\Lag^2_b \rpar \spc
\eq
\bqa
\Lag^0_b &=&
          - \lpar \mrF^0_{2\,\PH\PH} \rpar^2\,\mrA_0(M_{\PH}) 
          - \lpar \mrF^0_{2\,\PAz\PAz} \rpar^2\,\mrA_0(M_{\PAz}) 
\nl
{}&-& \frac{1}{2}\,\Bigl[
                 (2\,\mrF^0_{2\,\PHp\PHp}\,\mrF^0_{2\,\PHm\PHm} + 
                    \lpar \mrF^{00}_{2\,\PHp\PHm} \rpar^2 - 
                    \mrF^{00}_{2\,\PHp\PHm}\,\pdmu\,\mrF^{0\,\mu}_{2\,\PHp\PHm}
          \Bigr]\,\mrA_0(M_{\PHpm}) 
          - \frac{1}{2}\,\lpar \mrF^{00}_{2\,\PH\PAz} \rpar^2\,\mrB_{00}(\Msb) 
\nl
{}&-& \frac{1}{2}\,\Bigl[
                  2\,\lpar \mrF^0_{2\,\PH\PH} \rpar^2 + 2\,\lpar \mrF^0_{2\,\PAz\PAz} \rpar^2 + 
                  2\,\mrF^0_{2\,\PHp\PHp}\,\mrF^0_{2\,\PHm\PHm} + 
                  \lpar \mrF^{00}_{2\,\PHp\PHm} \rpar^2 - 
                  \mrF^{00}_{2\,\PHp\PHm}\,\pdmu\,\mrF^{0\,\mu}_{2\,\PHp\PHm}
                  \Bigr] \spc
\nl\nl
\Lag^2_b &=&
      - \frac{1}{4}\,\mrv^2\,g^2 \, \lpar \mrF^{00}_{2\,\PH\PAz} \rpar^2\,\lambda_5
\nl
{}&-& \frac{1}{12}\,g^2 \, \Bigl[
            24\,\mrF^0_{2\,\PH\PH}\,\mrF^2_{2\,\PH\PH} + 
            24\,\mrF^0_{2\,\PAz\PAz}\,\mrF^2_{2\,\PAz\PAz} + 
            12\,\mrF^0_{2\,\PHp\PHp}\,\mrF^2_{2\,\PHm\PHm} + 
            12\,\mrF^2_{2\,\PHp\PHp}\,\mrF^0_{2\,\PHm\PHm} 
\nl
{}&+&       12\,\mrF^2_{2\,\PHp\PHm}\,\mrF^{00}_{2\,\PHp\PHm} - 
             6\,\mrF^2_{2\,\PHp\PHm}\,\pdmu\,\mrF^{0\,\mu}_{2\,\PHp\PHm} - 
     \pdmu\pdnu\,\mrF^{0\,\nu}_{2\,\PHp\PHm}\,\pumu\,\mrF^{00}_{2\,\PHp\PHm} 
\nl
{}&+& 2\,\pdmu\,\mrF^0_{2\,\PHp\PHp}\,\pumu\,\mrF^0_{2\,\PHm\PHm} +
      2\,\pdmu\,\mrF^0_{2\,\PH\PH}\,\pumu\,\mrF^0_{2\,\PH\PH} 
\nl
{}&+& 3\,\pdmu\,\mrF^{00}_{2\,\PH\PAz}\,\pumu\,\mrF^{00}_{2\,\PH\PAz} +
      2\,\pdmu\,\mrF^0_{2\,\PAz\PAz}\,\pumu\,\mrF^0_{2\,\PAz\PAz} +
      \pdmu\,\mrF^{00}_{2\,\PHp\PHm}\,\pumu\,\mrF^{00}_{2\,\PHp\PHm} 
         \Bigr]
\nl
{}&-& 2\,g^2\,\mrF^0_{2\,\PH\PH}\,\mrF^2_{2\,\PH\PH}\,\mrA_0(M_{\PH})
          - 2\,g^2\,\mrF^0_{2\,\PAz\PAz}\,\mrF^2_{2\,\PAz\PAz}\,\mrA_0(M_{\PAz})
\nl
{}&-& \frac{1}{2}\,g^2\,\Bigl[
            2\,\mrF^0_{2\,\PHp\PHp}\,\mrF^2_{2\,\PHm\PHm} + 
            2\,\mrF^2_{2\,\PHp\PHp}\,\mrF^0_{2\,\PHm\PHm} 
\nl
{}&+&       2\,\mrF^2_{2\,\PHp\PHm}\,\mrF^{00}_{2\,\PHp\PHm} - 
            \mrF^2_{2\,\PHp\PHm}\,\pdmu\,\mrF^{0\,\mu}_{2\,\PHp\PHm}
          \Bigr]\,\mrA_0(M_{\PHpm})
          - g^2\,\mrF^2_{2\,\PH\PAz}\,\mrF^{00}_{2\,\PH\PAz}\,\mrB_{00}(\Msb) \spp
\eqa
\section{Conclusions \label{Concu}}
In this work we have been mainly interested in the effect of heavy scalars, with masses that 
are larger than the Higgs VEV and the energies probed by current experimental data. In particular
we focused on models where there are mixing effects in the mass matrices.
Therefore, we have adopted a top-down approach where there is a model and the aim has been to 
implement a systematic procedure for getting the low-energy theory, including all loop generated 
local operators.

We have extended the covariant derivative expansion~\cite{Henning:2014wua,Drozd:2015rsp}, 
taking into account SM extensions where heavy fields are coupled to (light) SM fields
with linear (or higher) couplings that are proportional to the scale of new physics.
Specific examples have been provided for the singlet extension of the SM and for THDM 
(I, II,X and Y) models~\cite{Yagyu:2012qp}. Working in the broken phase, including all 
contributions and normalizing the kinetic terms considerably increases the number of SM 
deviations as compared to what is usually reported in the literature.
\section{Acknowledgments}
We gratefully acknowledge several important discussions with Stefan~Dittmaier as well as his 
contribution in an early stage of this work. 
M.~B. and R.~G.~A. acknowledge discussions with Michael Spira during the HiggsTools Second 
Young Researchers Meeting. 
R.~G.~A. would like to thank Stefan~Dittmaier for the hospitality at Freiburg University where 
part of this work was done.

\section{Appendix: Expansion of loop integrals \label{eC0}}
Power counting of loop integrals can be summarized as follows: define
\bq
\int d^nq \frac{q_{\mu_1}\,\cdots\,q_{\mu_{2k}}}{(q^2 + M^2)^l} =
\delta_{\mu_1\,\dots\,\mu_{2k}}\,\mrI_{l\,,\,k} \spc
\eq
where the $\delta$ is the fully symmetric combination. In the large $M$ limit one has
\bq
\mrI_{1\,,\,2\,k} \sim M^{2 + 2\,k}\,\ln M^2 \spc
\quad
\mrI_{l\,,\,0} \sim (M^2)^{2-l}\;\; \spc
\mrI_{2\,,\,0} \sim \ln M^2 \spc
\quad
\mrI_{l\,,\,2\,k} \sim \mrI_{l - 1\,,\,2\,k-2} \;\; l > 1 \spp
\label{PCLI}
\eq
We define the following functions:
\bqa
i\,\pi^2\,\mrC^{(1)}_0(m) &=&
\muR^{4-n}\,\int d^n q\,
\frac{1}{(q^2 + m^2)\,((q+p_1)^2 + M^2_{\PH})\,((q+p_1+p_2)^2 + m^2)} \spc
\nl
i\,\pi^2\,\mrC^{(2)}_0(m) &=&
\muR^{4-n}\,\int d^n q\,
\frac{1}{(q^2 + m^2)\,((q+p_1)^2 + M^2_{\PH})\,((q+p_1+p_2)^2 + M^2_{\PH})} \spc
\eqa
with $P = p_1 + p_2$, $\muR$ being the renormalization scale. Their $M_{\PH}$ expansion is given 
in terms of two-point functions
\bq
i\,\pi^2\,\mrB_0\lpar \alpha\,,\,\beta\,;\,P^2\,,m\,,\,m\rpar =
\muR^{4-n}\,\int d^n q\,\frac{1}{(q^2 + m^2)^{\alpha}\,((q+P)^2 + m^2)^{\beta}} \spc
\eq
and of one-point functions, defined in \eqn{A0def}.
We obtain
\bqa
\mrC^{(2)}_0(m) &=& \frac{1}{M^2_{\PH}} + \frac{1}{M^4_{\PH}}\,\Bigl\{
   m^2\,\Bigl[ 1 - \mrA_0(M_{\PH}) + \mrA_0(m) \Bigr]
   - \frac{1}{3}\,\lpar p^2_1 + p^2_2 + \frac{1}{2}\,\spro{p_1}{p_2} \rpar \Bigr\} \spc
\nl
\mrC^{(2)}_0(m) &=& \frac{1}{M^2_{\PH}}\,\Bigl[ 
\mrB_0\lpar 1\,,\,1\,;\,P^2\,,\,m\,,\,m\rpar +
\mrA_0(m) + \ln \frac{M^2_{\PH}}{m^2} \Bigr]
\nl
{}&+& \frac{1}{M^4_{\PH}}\,\Bigl\{
     \frac{3}{4}\,\lpar p^2_1 + p^2_2 - \frac{1}{3}\,P^2 \rpar
   - \frac{1}{2}\,\lpar p^2_1 + p^2_2 - P^2 - 4\,m^2 \rpar\,\ln \frac{M^2_{\PH}}{m^2}
\nl
{}&+& \Bigl[ m^2 + \frac{1}{2}\,\lpar P^2 - p^2_1 - p^2_2 \rpar \Bigr]\,\mrA_0(m)
    + \Bigl[ m^2 + \frac{1}{2}\,\lpar P^2 - p^2_1 - p^2_2 \rpar \Bigr]\,
            \mrB_0\lpar 1\,,\,1\,;\,P^2\,,\,m\,,\,m\rpar \Bigr\} \spc
\label{Cexpa}
\eqa
where the $\mrB_0$ function is given by
\bq
\mrB_0\lpar 1\,,\,1\,;\,P^2\,,\,m\,,\,m\rpar  =
\frac{2}{4 - n} - \gamma - \ln \pi + 2 - \ln \frac{m^2}{\muRs} -
\beta\,\ln\frac{\beta + 1}{\beta - 1} \spc
\eq
with $\beta^2 = 1 + 4\,m^2/(P^2 - i\,0)$.

In deriving the expansion in \eqn{Cexpa} we also need the following results:
\bqa
{}&{}& \mrB^{\fin}_0\lpar 1\,,\,1\,;\,P^2\,,\,M_{\PH}\,,\,m\rpar  =
   1 - \ln\frac{M^2_{\PH}}{s} +
   \frac{s}{M^2_{\PH}}\,\lpar \frac{1}{2} - \frac{m^2}{s},\ln\frac{M^2_{\PH}}{m^2} \rpar 
\nl
{}&\qquad +&
   \frac{s^2}{M^4_{\PH}}\,\Bigl[ \frac{1}{6} + \frac{3}{2}\,\frac{m^2}{s} -
   \frac{m^2}{s}\,\lpar 1 + \frac{m^2}{s} \rpar\,\ln\frac{M^2_{\PH}}{m^2} \Bigr] \spc
\nl
{}&{}& \mrB^{\fin}_0\lpar 2\,,\,1\,;\,P^2\,,\,M_{\PH}\,,\,m\rpar  =
     \frac{s}{M^2_{\PH}} + \frac{s^2}{M^4_{\PH}}\,\Bigl[
          \frac{1}{2} + \frac{m^2}{s}\,\lpar 1 - \ln\frac{M^2_{\PH}}{m^2} \rpar \Bigr] \spp
\eqa
When all masses are heavy we derive:
\bq
\mrB_0\lpar 1\,,\,1\,;\,P^2\,,\,M_{\PH}\,,\,M_{\PH}\rpar =
 - \mrA_0(M_{\PH}) - 1 - \frac{1}{6}\,\frac{P^2}{M^2_{\PH}} + 
   \mcO\lpar\frac{P^4}{M^4_{\PH}} \rpar \spc
\eq
for the singlet extension and
\bq
\mrB_0\lpar 1\,,\,1\,;\,P^2\,,\,M_1\,,\,M_1\rpar =
 - \mrB_{00}\lpar M_1\,,\,M_2 \rpar + \frac{\mrv^2}{\Msbs}\mrB_{02}\lpar M_1\,,\,M_2 \rpar + 
   \frac{P^2}{\Msbs}\mrB_{0p}\lpar M_1\,,\,M_2 \rpar + \;\cdots \spc
\eq
for THDM models. Using the masses of \eqn{THDMexp} we obtain
\bqa
\mrB_{00}\lpar M_1\,,\,M_2 \rpar &=& - \mrB_{00}\lpar \Msb \rpar \spc
\qquad
\mrB_{0p}\lpar M_1\,,\,M_2 \rpar = - \frac{1}{2} \spc
\nl
\mrB_{02}\lpar M_{\PAz}\,,\,M_{\PHpm} \rpar &=& 
  - \frac{1}{4}\,\lpar \lambda_4 + 3\,\lambda_5 \rpar \spc
\quad
\mrB_{02}\lpar M_{\PH}\,,\,M_{\PHpm} \rpar =  - \frac{1}{4}\,\lpar \lambda_4 + \lambda_5 \rpar \spc
\quad
\mrB_{02}\lpar M_{\PH}\,,\,M_{\PAz} \rpar = - \frac{1}{2}\,\lambda_5 \spp
\eqa
\section{Appendix: Complete SESM Lagrangian \label{Lcomplete}}
In this Appendix we give the list of local operators, up to $\mcO(\Lambda^{-2})$, present in the
singlet extension of the SM after integration of the heavy mode $\PH$.
Field content is: gauge bosons $\PA\, \PZ\, \PWpm$, light Higgs $\Ph$, Higgs-Kibble ghosts
$\Ppz, \Pppm$, fermions $\PQu, \PQd$ and Faddeev-Popov ghosts $\PXp, \PXm, \PYa, \PYz$. 
Few auxiliary quantities are needed:
\bq
\PKz = \upphi^{\dagger}_0\,\upphi_{_0} = \Ppz^2 + 2\,\Ppp\,\Ppm \spc
\qquad
\PKh = \upphi^{\dagger}_{\Ph}\,\upphi_{_{\Ph}} = \Ph^2 + \Ppz^2 + 2\,\Ppp\,\Ppm \spp
\label{notI}
\eq
We also need $U(1)$ covariant derivatives:
\bq
\CDmu\,\Pppm = \pdmu\,\Pppm \pm i\,g\,\stW\,\PA_{\mu}\,\Pppm \spc
\qquad
\CDmu\,\Pf = \pdmu\,\Pf - i\,g\,\mrQ_{\Pf}\,\stW\,\PA_{\mu}\,\Pf \spp
\eq
Finally, $\mrT$ functions are defined in \eqn{Tdef}, while $\beta$ coefficients in 
\eqn{Bdef}; the one-point function $\mrA_0$ is defined in \eqn{A0def}. Dimension and codimension 
of local operators are defined in \eqn{dimdef}. In the following list we give 
$\Lag_{\mrdim\,,\,\cdim}$ for one flavor, \ie
\bq
\Lag = \sum_{n=0}^{2}\,\Lambda^{2\,n - 2}\,\sum_l\,M^l\,\Lag_{6-2\,n\,,\,6-2\,n-l} \spp
\eq

\bei
\item {\underline{$\mrdim = 2$}}
\bqa
\blacktriangleright\;\Lag_{22} &=&
       -
         \frac{1}{32}
         \,\frac{g^2}{\pi^2}
         \,\PKh
         \,\beta^{(0}_2 -
         \frac{1}{16}
         \,\frac{g^2}{\pi^2}
         \,t_m\,t^2_1
         \,\PKz
         \,\mrA_0 \spc
\quad
\Lag_{23} =
         \frac{1}{32}
         \,\frac{g^3}{\pi^2}
         \,\frac{t_m\,t^2_1}{M}
         \,\PKh
         \,\Ph
         \,\mrA_0 \spc
\quad
\Lag_{24} =
         \frac{1}{256}
         \,\frac{g^4}{\pi^2}
         \,\frac{t_m\,t^2_1}{M^2}
         \,\PKhs
         \,\mrA_0 \spc
\nl
\eqa
\item {\underline{$\mrdim = 4$}}
\bqa
\blacktriangleright\;\Lag_{41} &=&
       -
         \frac{1}{8}
         \,\frac{g\,M^3}{\pi^2}
         \,\Ph
         \,\Delta \beta^{(1)}_2 \spc
\eqa
\bqa
\blacktriangleright\;\Lag_{42} &=&
       -
         \frac{1}{32}
         \,\frac{g^2\,M^2}{\pi^2}
         \,\PKh
         \,\beta^{(1)}_2
       +
         \frac{1}{32}
         \,\frac{g^2\,M^2}{\pi^2}
         \,t^2_m
         \,\lpar  \beta^{(0)}_2 - \beta^{(0)}_1 \rpar
         \,\Ph^2
\nl {}&+&
         \frac{1}{16}
         \,\frac{g^2\,M^2}{\pi^2}
         \,t^2_m\,t_1
         \,\lpar \PAXm\,\PXm + \PAXp\,\PXp \rpar
         \,\mrA_0
       -
         \frac{1}{64}
         \,\frac{g^2\,M^2}{\pi^2}
         \,t^2_m\,t_1
         \,\lpar x_{\Ph} + 4\,t^2_m\,t_1 \rpar
         \,\Ppz^2
         \,\mrA_0
\nl {}&-&
         \frac{1}{32}
         \,\frac{g^2\,M^2}{\pi^2}
         \,t^2_m\,t_1
         \,\Bigl[ x_{\Ph} - 2\,\lpar 1 - 2\,t^2_m\,t_1 \rpar \Bigr]
         \,\Ppp
         \,\Ppm
         \,\mrA_0
      +
         \frac{1}{32}
         \,\frac{g^2\,M^2}{\pi^2}
         \,t_m
         \,\PKh
         \,\mrT_1
\nl {}&+&
         \frac{1}{32}
         \,\frac{g^2\,M^2}{\ctWs\,\pi^2}
         \,t^2_m\,t_1
         \,\lpar \Ppz^2 + 2\,\PAYz\,\PYz \rpar
         \,\mrA_0
       -
         \frac{1}{2}
         \,\frac{M^2}{\ctWs}
         \,\lpar \Ppz^2 + 2\,\PAYz\,\PYz \rpar
\nl {}&-&
         \frac{1}{2}
         \,M^2
         \,\Bigl[ \lpar 2\,\PWpdmu\,\PWmdmu + \frac{1}{\ctWs}\,\PZdmus \rpar + 
                       2\,\lpar \PAXm\,\PXm + \PAXp\,\PXp \rpar + 2\,\Ppp\,\Ppm \Bigr]
\nl {}&-&
         \frac{1}{2}
         \,M^2\,x_{\Ph}
         \,\Ph^2 \spc
\eqa
\bqa
\blacktriangleright\;\Lag_{43} &=&
         -\,\lpar  \PdAXmmu\,\PdXmmu + \PdAXpmu\,\PdXpmu + \PdAYzmu\,\PdYzmu + 
                  \PdAYamu\,\PdYamu \rpar
\nl {}&+&
         \frac{1}{64}
         \,\frac{g^3\,M}{\pi^2}
         \,t^2_m
         \,\lpar  \beta^{(0)}_2 - \beta^{(0)}_1 \rpar
         \,\PKh
         \,\Ph
\nl {}&+&
         \frac{1}{64}
         \,\frac{g^3\,M}{\pi^2}
         \,t^2_m\,t_1
         \,\Bigl[ \lpar 2\,\PWpdmu\,\PWmdmu + \frac{1}{\ctWs}\,\PZdmus \rpar + 
                 \lpar \PAXm\,\PXm + \PAXp\,\PXp \rpar \Bigr]
         \,\Ph
         \,\mrA_0
\nl {}&+&
         \frac{1}{64}
         \,\frac{g^3\,M}{\pi^2}
         \,t^2_m\,t_1
         \,\Bigl[ x_{\Ph}\,\mrA_0 - 8\,\lpar 1 - t_m \rpar\,t_m\,t_1 - 
                 \lpar 13 - 14\,t_m \rpar\,t_m\,t_1\,\mrA_0 \Bigr]
         \,\PKz
         \,\Ph
\nl {}&+&
         \frac{1}{64}
         \,\frac{g^3\,M}{\pi^2}
         \,t^2_m\,t_1
         \,\Bigl[ x_{\Ph}\,\mrA_0 - 8\,\lpar 1 - t_m \rpar\,t_m\,t_1 - 
                 \lpar 15 - 14\,t_m \rpar\,t_m\,t_1\,\mrA_0 \Bigr]
         \,\Ph^3
\nl {}&+&
         \frac{1}{64}
         \,\frac{g^3\,M}{\ctWs\,\pi^2}
         \,\PAYz
         \,\PYz
         \,t^2_m\,t_1
         \,\Ph
         \,\mrA_0
       +
         \frac{1}{2}
         \,\frac{i\,g\,M}{\ctW}
         \,\lpar 1 - 2\,\ctWs \rpar
         \,\lpar \PAXp\,\Ppp - \PAXm\,\Ppm  \rpar
         \,\PYz
\nl {}&+&
         \frac{1}{2}
         \,\frac{i\,g\,M}{\ctW}
         \,\PAYz
         \,\lpar \PXm\,\Ppp - \PXp\,\Ppm \rpar
\nl {}&-&
         \frac{1}{64}
         \,\frac{i\,g^3\,M}{\pi^2}
         \,t^2_m\,t_1
         \,\Bigl[ 2\,\lpar \Ppp\,\PWmmu - \Ppm\,\PWpmu \rpar\,
                     \lpar  \frac{\stWs}{\ctW}\,\PZmu - \PAmu\,\stW  \rpar + 
                     \lpar  \PAXp\,\PXp  - \PAXm\,\PXm\rpar\,\Ppz  \Bigr]
         \,\mrA_0
\nl {}&-&
         \frac{1}{64}
         \,\frac{i\,g^3\,M}{\pi^2}
         \,\lpar 1 - 2\,\ctWs \rpar
         \,\frac{1}{\ctW\,t^2_m\,t_1,t^2_m\,t_1}
         \,\lpar  \PAXp\,\Ppp - \PAXm\,\Ppm  \rpar
         \,\PYz
         \,\mrA_0
\nl {}&-&
         \frac{1}{64}
         \,\frac{i\,g^3\,M}{\pi^2}
         \,\frac{1}{\ctW\,t^2_m\,t_1\,t^2_m\,t_1}
         \,\PAYz
         \,\lpar \PXm\,\Ppp - \PXp\,\Ppm \rpar
         \,\mrA_0
\nl {}&+&
         \frac{1}{32}
         \,\frac{i\,g^3\,M}{\pi^2}
         \,t^2_m\,t_1\,\stW
         \,\lpar \PAXp\,\Ppp - \PAXm\,\Ppm  \rpar
         \,\PYa
         \,\mrA_0
\nl {}&-&
         \frac{1}{2}
         \,\frac{g\,M}{\ctWs}
         \,\PAYz
         \,\PYz
         \,\Ph
       -
         \frac{1}{4}
         \,g\,M\,x_{\Ph}
         \,\PKh
         \,\Ph
\nl {}&+&
         \frac{1}{2}
         \,i\,g\,M
         \,\Bigl[ 2\,\lpar \Ppp\,\PWmmu - \Ppm\,\PWpmu \rpar\,
                     \lpar  \frac{\stWs}{\ctW}\,\PZmu - \PAmu\,\stW  \rpar + 
                     \lpar  \PAXp\,\PXp  - \PAXm\,\PXm \rpar\,\Ppz  \Bigr]
\nl {}&-&
         \frac{1}{2}
         \,g\,M
         \,\Bigl[ \lpar 2\,\PWpdmu\,\PWmdmu + \frac{1}{\ctWs}\,\PZdmus \rpar + 
                 \lpar \PAXm\,\PXm + \PAXp\,\PXp \rpar \Bigr]
         \,\Ph
\nl {}&-&
         M\,x_{\PQd}
         \,\PAQd\,\PQd
       -
         M\,x_{\PQu}
         \,\PAQu\,\PQu
       -
         i\,g\,M
         \,\stW
         \,\lpar  \PAXp\,\Ppp - \PAXm\,\Ppm  \rpar
         \,\PYa \spc
\eqa
\bqa
\blacktriangleright\;\Lag_{44} &=&
       -
         \frac{1}{2}
         \,\Bigl[ \PkinA + \PkinZ + 2\,\PkinW + \pdmuPh^2 + \pdmuPpz^2 
\nl {}&+& 
           2\,\PAQu\,\gamma^{\mu}\,\PDQu + 2\,\PAQd\,\gamma^{\mu}\,\PDQd \Bigr]
       -
         \PdmuPppm
       -
         \frac{1}{2}
         \,\frac{g^2\,\stWs}{\ctW}
         \,\lpar \Ppp\,\PWmmu + \Ppm\,\PWpmu \rpar
         \,\Ppz
         \,\PZmu
\nl {}&-&
         \frac{1}{64}
         \,\frac{g^3}{\pi^2}
         \,t^2_m\,t_1\,\lpar x_{\PQd}\,\PAQd\,\PQd + x_{\PQu}\,\PAQu\,\PQu \rpar
         \,\Ph
         \,\mrA_0
       -
         \frac{1}{512}
         \,\frac{g^4}{\pi^2}
         \,t^2_m
         \,\PKhs
         \,\beta^{(0)}_1
\nl {}&+&
         \frac{1}{1024}
         \,\frac{g^4}{\pi^2}
         \,t^2_m\,t_1
         \,\Bigl[ x_{\Ph}\,\mrA_0 - 16\,\lpar 2 - t_m \rpar\,t_m\,t_1 - 
                 28\,\lpar 2 - t_m \rpar\,t_m\,t_1\,\mrA_0 \Bigr]
         \,\PKzs
\nl {}&+&
         \frac{1}{512}
         \,\frac{g^4}{\pi^2}
         \,t^2_m\,t_1
         \,\Bigl[ x_{\Ph}\,\mrA_0 - 16\,\lpar 2 - t_m \rpar\,t_m\,t_1 - 
                 4\,\lpar 17 - 7\,t_m \rpar\,t_m\,t_1\,\mrA_0 \Bigr]
         \,\PKz
         \,\Ph^2
\nl {}&+&
         \frac{1}{1024}
         \,\frac{g^4}{\pi^2}
         \,t^2_m\,t_1
         \,\Bigl[ x_{\Ph}\,\mrA_0 - 
                  16\,\lpar 2 - t_m \rpar\,t_m\,t_1 - 
                   4\,\lpar 20 - 7\,t_m \rpar\,t_m\,t_1\,\mrA_0 \Bigr]
         \,\Ph^4
\nl {}&+&
         \frac{1}{4}
         \,\frac{i\,g}{\ctW}
         \,\lpar \PAQu\,\gamma^{\mu}\,\PQu + \PAQu\,\gamma^{\mu}\,\gamma^5\,\PQu \rpar
         \,\PZmu
\nl {}&+&
         \frac{1}{4}
         \,\frac{i\,g}{\ctW}
         \,\lpar 1 - 2\,\ctWs \rpar
         \,\lpar \PAQd\,\gamma^{\mu}\,\PQd + \PAQd\,\gamma^{\mu}\,\gamma^5\,\PQd \rpar
         \,\PZmu
\nl {}&-&
         \frac{1}{2}
         \,\frac{i\,g}{\ctW}
         \,\lpar 1 - 2\,\ctWs \rpar
         \,\Ppp 
         \,\PZmu
         \,\PdmuPpm
       +
         \frac{1}{2}
         \,\frac{i\,g}{\ctW}
         \,\lpar 1 - 2\,\ctWs \rpar
         \,\Ppm 
         \,\PZmu
         \,\PdmuPpp
\nl {}&-&
         \frac{i\,g\,\stWs}{\ctW}
         \,\lpar \mrQ_{\PQu}\,\PAQu\,\gamma^{\mu}\,\PQu +
                 \mrQ_{\PQd}\,\PAQd\,\gamma^{\mu}\,\PQd \rpar 
         \,\PZmu
       -
         \frac{1}{2}
         \,\frac{i\,g\,\stWs}{\ctW}
         \,\lpar \PAQd\,\gamma^{\mu}\,\PQd + \PAQd\,\gamma^{\mu}\,\gamma^5\,\PQd \rpar
         \,\PZmu
\nl {}&+&
         \frac{1}{2}
         \,\frac{i\,g^2\,\stWs}{\ctW}
         \,\lpar \Ppp\,\PWmmu - \Ppm\,\PWpmu \rpar
         \,\Ph
         \,\PZmu
\nl {}&-&
         \frac{1}{64}
         \,\frac{i\,g^3}{\pi^2}
         \,t^2_m\,t_1\,\lpar
          x_{\PQd}\,\PAQd\,\gamma^5\,\PQd +
          x_{\PQu}\,\PAQu\,\gamma^5\,\PQu \rpar
         \,\Ppz
         \,\mrA_0
\nl {}&-&
         \frac{1}{64}
         \,\frac{i\,g^3}{\srt\,\pi^2}
         \,t^2_m\,t_1\,x_{\PQd}
         \,\lpar \PAQu\,\gamma_{-}\,\Ppp\,\PQd - \PAQd\,\gamma_{+}\,\Ppm\,\PQu \rpar
         \,\mrA_0
\nl &+&
         \frac{1}{64}
         \,\frac{i\,g^3}{\srt\,\pi^2}
         \,t^2_m\,t_1\,x_{\PQu}
         \,\lpar \PAQu\,\gamma_{+}\,\Ppp\,\PQd - \PAQd\,\gamma_{-}\,\Ppm\,\PQu \rpar
         \,\mrA_0
\nl {}&+&
         \frac{1}{2}
         \,\frac{i\,g}{\srt}
         \,\lpar \PAQu\,\gamma^{\mu}\,\gamma_{+}\,\PQd\,\PWpmu + 
                \PAQd\,\gamma^{\mu}\,\gamma_{+}\,\PQu\,\PWmmu \rpar
       +
         \frac{1}{2}
         \,\frac{i}{\srt}
         \,g\,x_{\PQu}
         \,\lpar \PAQu\,\gamma_{+}\,\Ppp\,\PQd - \PAQd\,\gamma_{-}\,\Ppm\,\PQu \rpar
\nl {}&-&
         \frac{1}{2}
         \,\frac{i}{\srt}
         \,g\,x_{\PQd}
         \,\lpar \PAQu\,\gamma_{-}\,\Ppp\,\PQd - \PAQd\,\gamma_{+}\,\Ppm\,\PQu \rpar
\nl {}&-&
         \frac{1}{2}
         \,\frac{g}{\ctW}
         \,\lpar  \Ppz\,\pdmuPh - \Ph\,\pdmuPpz  \rpar
         \,\PZmu
\nl {}&-&
         \frac{1}{32}
         \,g^2\,x_{\Ph}
         \,\PKhs
       +
         g^2
         \,\stWs
         \,\lpar \Ppp\,\Ppm \rpar
         \,\PZdmus
\nl {}&-&
         \frac{1}{8}
         \,g^2
         \,\lpar 2\,\PWpdmu\,\PWmdmu + \frac{1}{\ctWs}\,\PZdmus \rpar
         \,\PKh
\nl {}&+&
         \frac{1}{2}
         \,i\,g\,\lpar x_{\PQu}\,\PAQu\,\gamma^5\,\PQu +
                       x_{\PQd}\,\PAQd\,\gamma^5\,\PQd\rpar
         \,\Ppz
\nl {}&-&
         \frac{1}{2}
         \,i\,g
         \,\PWpmu
         \,\Ppz
         \,\PdmuPpm
       +
         \frac{1}{2}
         \,i\,g
         \,\PWmmu
         \,\Ppz
         \,\PdmuPpp
\nl {}&-&
         \frac{1}{2}
         \,i\,g
         \,\Bigl[ \lpar \Ppp\,\PWmmu - \Ppm\,\PWpmu \rpar\,\pdmuPpz - 
               2\,\lpar \PAmu\,\stW + \PZmu\,\ctW \rpar\,\PdWpnumu\,\PWmnu 
\nl {}&+& 
               2\,\lpar \PAmu\,\stW + \PZmu\,\ctW \rpar\,
                 \PdWmnumu\,\PWpnu + 2\,\lpar \PAXp\,\PXp - \PAXm\,\PXm  \rpar\,
                  \lpar \PdivA\,\stW + \PdivZ\,\ctW \rpar 
\nl {}&+& 2\,\lpar  
                 \PAXp\,\PdXpmu 
                 - \PAXm\,\PdXmmu  
                 \rpar\,
                 \lpar \PAmu\,\stW + \PZmu\,\ctW \rpar \Bigr]
\nl {}&+&
         i\,g\,\ctW
         \,\Bigl[ \lpar \PWpmu\,\PdWmnumu - \PWmmu\,\PdWpnumu \rpar\,\PZnu - 
                 \lpar \PWpmu\,\PWmnu - \PWmmu\,\PWpnu \rpar\,\PdZnumu 
\nl {}&+& 
                 \lpar  \PAXp\,\PYz - \PAYz\,\PXm  \rpar\,\PdivWp + 
                 \lpar  \PAYz\,\PXp - \PAXm\,\PYz \rpar\,\PdivWm + 
                 \lpar  \PAXp\,\PdYzmu - \PAYz\,\PdXmmu  \rpar\,\PWpmu 
\nl {}&-& \lpar  \PAYz\,\PdXpmu + \PAXm\,\PdYzmu \rpar\,\PWmmu \Bigr]
\nl {}&+&
         i\,g\,\stW
         \,\Bigl[ \lpar \PWpmu\,\PdWmnumu - \PWmmu\,\PdWpnumu \rpar\,\PAnu - 
                 \lpar \PWpmu\,\PWmnu - \PWmmu\,\PWpnu \rpar\,\PdAnumu 
\nl {}&+& 
                 \lpar \PAXp\,\PYa - \PAYa\,\PXm  \rpar\,\PdivWp + 
                 \lpar \PAYa\,\PXp - \PAXm\,\PYa \rpar\,\PdivWm + 
                 \lpar \PAXp\,\PdYamu - \PAYa\,\PdXmmu \rpar\,\PWpmu 
\nl {}&-& 
                 \PAXm\,\PdYamu\,\PWmmu + \PAYa\,\PdXpmu\,\PWmmu \Bigr]
\nl {}&-&
         \frac{1}{2}
         \,g
         \,\Bigl[ \lpar \Ppp\,\PWmmu + \Ppm\,\PWpmu \rpar\,\pdmuPh - 
                  \lpar \PWpmu\,\PdmuPpm + \PWmmu\,\PdmuPpp \rpar\,\Ph \Bigr]
\nl {}&-&
         \frac{1}{2}
         \,g\,\lpar x_{\PQu}\,\PAQu\,\PQu +
                    x_{\PQd}\,\PAQd\,\PQd \rpar
         \,\Ph
\eqa
\bqa
\blacktriangleright\;\Lag_{45} &=&
       -
         \frac{1}{256}
         \,\frac{g^5}{\pi^2}
         \,\frac{t^3_m\,t^2_1}{M}
         \,\Ph
         \,\PKhs
         \,\mrA_0 \spc
\eqa
\item {\underline{$\mrdim = 6$}}
\bqa
\blacktriangleright\;\Lag_{61} &=&
       -
         \frac{1}{8}
         \,\frac{g\,M^5}{\pi^2}
         \,\Ph
         \,\Delta \beta^{(2)}_2 +
         \frac{1}{16}
         \,\frac{g\,M^5}{\pi^2}
         \,t^2_m
         \,\Ph
         \,\Delta \beta^{(1)}_2 \spc
\eqa
\bqa
\blacktriangleright\;\Lag_{62} &=&
       -
         \frac{1}{32}
         \,\frac{g^2\,M^4}{\pi^2}
         \,\PKh
         \,\beta^{(1)}_2
       +
         \frac{1}{64}
         \,\frac{g^2\,M^4}{\pi^2}
         \,\frac{t^2_m}{t_1}
         \,\lpar  \beta^{(0)}_2 - \beta^{(0)}_1 \rpar
         \,\lpar x_{\Ph} - 2\,t^2_m\,t_1 \rpar
         \,\Ph^2
\nl {}&+&
         \frac{1}{64}
         \,\frac{g^2\,M^4}{\pi^2}
         \,\frac{t_m}{t_1}
         \,\lpar 2\,\PWpdmu\,\PWmdmu + \frac{1}{\ctWs}\,\PZdmus \rpar
         \,\mrT_1
       -
         \frac{1}{64}
         \,\frac{g^2\,M^4}{\pi^2}
         \,\frac{t_m}{t_1}
         \,\lpar 3 + 2\,t^2_m\,t_1 \rpar
         \,\PKz
         \,\mrT_1
\nl {}&+&
         \frac{1}{64}
         \,\frac{g^2\,M^4}{\pi^2}
         \,\frac{t_m}{t_1}
         \,\Bigl[ x_{\Ph} - \lpar 3 - 2\,t_m\,t_1 + 4\,t^2_m\,t_1 \rpar \Bigr]
         \,\Ph^2
         \,\mrT_1
       +
         \frac{1}{32}
         \,\frac{g^2\,M^4}{\pi^2}
         \,t^2_m
         \,\PKz
         \,\Delta \beta^{(1)}_2
\nl {}&+&
         \frac{1}{32}
         \,\frac{g^2\,M^4}{\pi^2}
         \,t^2_m
         \,\lpar   \beta^{(1)}_2 - \beta^{(1)}_1  + \Delta \beta^{(1)}_2 \rpar
         \,\Ph^2
       +
         \frac{1}{32}
         \,\frac{g^2\,M^4}{\pi^2}
         \,t^2_m\,x_{\Ph}
         \,\lpar \PAXm\,\PXm + \PAXp\,\PXp \rpar
         \,\mrA_0
\nl {}&-&
         \frac{1}{128}
         \,\frac{g^2\,M^4}{\pi^2}
         \,t^2_m\,x_{\Ph}
         \,\Bigl[ x_{\Ph} - 2\,\lpar 1 - 3\,t_m \rpar\,t_m\,t_1 \Bigr]
         \,\Ppz^2
         \,\mrA_0
\nl {}&-&
         \frac{1}{64}
         \,\frac{g^2\,M^4}{\pi^2}
         \,t^2_m\,x_{\Ph}
         \,\Bigl[ x_{\Ph} - 2\,\lpar 1 + t_m\,t_1 - 3\,t^2_m\,t_1 \rpar \Bigr]
         \,\Ppp
         \,\Ppm
         \,\mrA_0
\nl {}&+&
         \frac{1}{32}
         \,\frac{g^2\,M^4}{\pi^2}
         \,t_m
         \,\PKh
         \,\mrT_2
       +
         \frac{1}{64}
         \,\frac{g^2\,M^4}{\ctWs\,\pi^2}
         \,t^2_m\,x_{\Ph}
         \,\lpar \Ppz^2 + 2\,\PAYz\,\PYz \rpar
         \,\mrA_0 \spc
\eqa
\bqa
\blacktriangleright\;\Lag_{63} &=&
       -
         \frac{1}{64}
         \,\frac{g^2\,M^3}{\pi^2}
         \,\frac{t_m}{t_1}
         \,\lpar \PWpmu\,\PdmuPpm + \PWmmu\,\PdmuPpp \rpar
         \,\mrT_1
\nl {}&+&
         \frac{1}{64}
         \,\frac{g^2\,M^3}{\pi^2}
         \,\frac{t_m}{t_1}
         \,\lpar x_{\PQu}\,\PAQu\,\PQu + x_{\PQd}\,\PAQd\,\PQd \rpar
         \,\mrT_1
        -
         \frac{1}{64}
         \,\frac{g^2\,M^3}{\ctW\,\pi^2}
         \,\frac{t_m}{t_1}
         \,\mrT_1
         \,\PZmu
         \,\pdmuPpz
\nl {}&+&
         \frac{1}{128}
         \,\frac{g^3\,M^3}{\pi^2}
         \,\frac{t^2_m}{t_1}
         \,\lpar  \beta^{(0)}_2 - \beta^{(0)}_1 \rpar
         \,\lpar 2\,\PWpdmu\,\PWmdmu + \frac{1}{\ctWs}\,\PZdmus \rpar
         \,\Ph
\nl {}&+&
         \frac{1}{128}
         \,\frac{g^3\,M^3}{\pi^2}
         \,\frac{t^2_m}{t_1}
         \,\lpar  \beta^{(0)}_2 - \beta^{(0)}_1 \rpar
         \,\lpar x_{\Ph} - 3\,t^2_m\,t_1 \rpar
         \,\PKz
         \,\Ph
\nl {}&+&
         \frac{1}{128}
         \,\frac{g^3\,M^3}{\pi^2}
         \,\frac{t^2_m}{t_1}
         \,\lpar  \beta^{(0)}_2 - \beta^{(0)}_1 \rpar
         \,\Bigl[ 2\,x_{\Ph} + \lpar 2 - 5\,t_m \rpar\,t_m\,t_1 \Bigr]
         \,\Ph^3
\nl {}&+&
         \frac{1}{128}
         \,\frac{g^3\,M^3}{\pi^2}
         \,\frac{t_m}{t_1}
         \,\lpar 2\,\PWpdmu\,\PWmdmu + \frac{1}{\ctWs}\,\PZdmus \rpar
         \,\Ph
         \,\mrT_1
\nl {}&+&
         \frac{1}{256}
         \,\frac{g^3\,M^3}{\pi^2}
         \,\frac{t_m}{t_1}
         \,\lpar x_{\Ph} + 4\,\lpar 1 - t_m \rpar\,t_m\,t_1 \rpar
         \,\PKh
         \,\Ph
         \,\mrT_1
\nl {}&-&
         \frac{1}{64}
         \,\frac{g^3\,M^3}{\pi^2}
         \,t^2_m
         \,\lpar \beta^{(1)}_1 - \beta^{(1)}_2 \rpar
         \,\PKh
         \,\Ph
\nl {}&+&
         \frac{1}{384}
         \,\frac{g^3\,M^3}{\pi^2}
         \,t^2_m
         \,\Bigl[ 3\,x_{\Ph}\,\mrA_0 + 2\,\lpar 1 - t_m \rpar^2\,t_1 \Bigr]
         \,\lpar \PAXm\,\PXm + \PAXp\,\PXp \rpar
         \,\Ph
\nl {}&+&
         \frac{1}{384}
         \,\frac{g^3\,M^3}{\pi^2}
         \,t^2_m
         \,\Bigl[ 3\,x_{\Ph}\,\mrA_0 - 2\,\lpar 1 - t_m \rpar\,
                  \lpar 17 - 5\,t_m \rpar\,t_1 - 3\,
                  \lpar 18 - 22\,t_m + 7\,t^2_m \rpar\,t_1\,\mrA_0 \Bigr]
\nl {}&\times&
         \,\lpar 2\,\PWpdmu\,\PWmdmu + \frac{1}{\ctWs}\,\PZdmus \rpar
         \,\Ph
\nl {}&+&
         \frac{1}{384}
         \,\frac{g^3\,M^3}{\pi^2}
         \,t^2_m
         \,\Bigl[ 3\,x^2_{\Ph}\,\mrA_0 + 24\,\lpar 1 - t_m \rpar\,t^3_m\,t^2_1 - 
                  \lpar 1 + 40\,t_m - 53\,t^2_m \rpar\,t_1\,x_{\Ph} + 
               6\,\lpar 6 - 7\,t_m \rpar\,t^3_m\,t^2_1\,\mrA_0 
\nl {}&-& 
               3\,\lpar 19 - 25\,t_m \rpar\,t_m\,t_1\,x_{\Ph}\,\mrA_0 \Bigr]
         \,\PKz
         \,\Ph
\nl {}&+&
         \frac{1}{384}
         \,\frac{g^3\,M^3}{\pi^2}
         \,t^2_m
         \,\Bigl[ 3\,x^2_{\Ph}\,\mrA_0 + 8\,\lpar 1 - t_m \rpar\,
                 \lpar 4 - 20\,t_m + 19\,t^2_m \rpar\,t_m\,t^2_1 
\nl {}&-&   6\,\lpar 6 + 10\,t_m - 42\,t^2_m + 29\,t^3_m \rpar\,t_m\,t^2_1\,\mrA_0 
\nl {}&-& 
               3\,\lpar 18 - 3\,t_m - 17\,t^2_m \rpar\,t_1\,x_{\Ph}\,\mrA_0 -
                 \lpar 35 - 4\,t_m - 43\,t^2_m \rpar\,t_1\,x_{\Ph} \Bigr]
         \,\Ph^3
\nl {}&+&
         \frac{1}{384}
         \,\frac{g^3\,M^3}{\ctWs\,\pi^2}
         \,\PAYz
         \,\PYz
         \,t^2_m
         \,\Bigl[ 3\,x_{\Ph}\,\mrA_0 + 2\,\lpar 1 - t_m \rpar^2\,t_1 \Bigr]
         \,\Ph
\nl {}&-&
         \frac{1}{64}
         \,\frac{i\,g^3\,M^3}{\pi^2}
         \,\frac{t_m}{t_1}
         \,\frac{\stWs}{\ctW}
         \,\lpar \Ppp\,\PWmmu - \Ppm\,\PWpmu \rpar
         \,\mrT_1
         \,\PZmu
\nl {}&-&
         \frac{1}{128}
         \,\frac{i\,g^3\,M^3}{\pi^2}
         \,t^2_m\,x_{\Ph}
         \,\Bigl[ 2\,\lpar \Ppp\,\PWmmu - \Ppm\,\PWpmu \rpar\,
                 \lpar \frac{\stWs}{\ctW}\,\PZmu - \PAmu\,\stW  \rpar + 
                 \lpar \PAXp\,\PXp - \PAXm\,\PXm  \rpar\,\Ppz \Bigr]
         \,\mrA_0
\nl {}&-&
         \frac{1}{128}
         \,\frac{i\,g^3\,M^3}{\pi^2}
         \,\lpar 1 - 2\,\ctWs \rpar
         \,\frac{1}{\ctW\,t^2_m\,x_{\Ph},t^2_m\,x_{\Ph}}
         \,\lpar \PAXp\,\Ppp - \PAXm\,\Ppm  \rpar
         \,\PYz
         \,\mrA_0
\nl {}&-&
         \frac{1}{128}
         \,\frac{i\,g^3\,M^3}{\pi^2}
         \,\frac{1}{\ctW\,t^2_m\,x_{\Ph},t^2_m\,x_{\Ph}}
         \,\PAYz
         \,\lpar \PXm\,\Ppp - \PXp\,\Ppm \rpar
         \,\mrA_0
\nl {}&+&
         \frac{1}{64}
         \,\frac{i\,g^3\,M^3}{\pi^2}
         \,\stW
         \,t^2_m\,x_{\Ph}
         \,\lpar \PAXp\,\Ppp - \PAXm\,\Ppm  \rpar
         \,\PYa
         \,\mrA_0
\nl {}&+&
         \frac{1}{4}
         \,g\,M^3\,t^2_m
         \,\lpar 2\,\PWpdmu\,\PWmdmu + \frac{1}{\ctWs}\,\PZdmus \rpar
         \,\Ph
       +
         \frac{1}{8}
         \,g\,M^3\,t^2_m\,x_{\Ph}
         \,\lpar \PKh + 2\,\Ph^2 \rpar
         \,\Ph \spc
\eqa
\bqa
\blacktriangleright\;\Lag_{64} &=&
       -
         \frac{1}{128}
         \,\frac{g^3\,M^2}{\pi^2}
         \,\frac{t^2_m}{t_1}
         \,\lpar  \beta^{(0)}_2 - \beta^{(0)}_1 \rpar
         \,\lpar \PWpmu\,\PdmuPpm + \PWmmu\,\PdmuPpp \rpar
         \,\Ph
\nl {}&+&
         \frac{1}{128}
         \,\frac{g^3\,M^2}{\pi^2}
         \,\frac{t^2_m}{t_1}
         \,\lpar  \beta^{(0)}_2 - \beta^{(0)}_1 \rpar
         \,\lpar x_{\PQu}\,\PAQu\,\PQu + x_{\PQd}\,\PAQd\,\PQd \rpar
         \,\Ph
\nl {}&-&
         \frac{1}{384}
         \,\frac{g^3\,M^2}{\pi^2}
         \,t^2_m\,x_{\PQd}
         \,\Bigl[ 3\,x_{\Ph}\,\mrA_0 + 2\,\lpar 1 - t_m \rpar\,
                  \lpar 17 + t_m \rpar\,t_1 + 3\,
                  \lpar 18 - 18\,t_m + t^2_m \rpar\,t_1\,\mrA_0 \Bigr]
         \,\PAQd\,\PQd
         \,\Ph
\nl {}&-&
         \frac{1}{384}
         \,\frac{g^3\,M^2}{\pi^2}
         \,t^2_m\,x_{\PQu}
         \,\Bigl[ 3\,x_{\Ph}\,\mrA_0 + 2\,
                  \lpar 1 - t_m \rpar\,\lpar 17 + t_m \rpar\,t_1 + 3\,
                  \lpar 18 - 18\,t_m + t^2_m \rpar\,t_1\,\mrA_0 \Bigr]
         \,\PAQu\,\PQu
         \,\Ph
\nl {}&-&
         \frac{1}{192}
         \,\frac{g^3\,M^2}{\pi^2}
         \,t^2_m\,t_1
         \,\Bigl[ 3\,t^2_m\,\mrA_0 - \lpar 1 - t_m \rpar^2 \Bigr]
         \,\lpar \Ppp\,\PWmmu + \Ppm\,\PWpmu \rpar
         \,\pdmuPh
\nl {}&+&
         \frac{1}{192}
         \,\frac{g^3\,M^2}{\pi^2}
         \,t^2_m\,t_1
         \,\Bigl[ \lpar 1 - t_m \rpar\,\lpar 17 + t_m \rpar + 
                  3\,\lpar 9 - 9\,t_m + t^2_m \rpar\,\mrA_0 \Bigr]
         \,\lpar \PWpmu\,\PdmuPpm + \PWmmu\,\PdmuPpp \rpar
         \,\Ph
\nl {}&-&
         \frac{1}{128}
         \,\frac{g^3\,M^2}{\ctW\,\pi^2}
         \,\frac{t^2_m}{t_1}
         \,\lpar  \beta^{(0)}_2 - \beta^{(0)}_1 \rpar
         \,\Ph
         \,\PZmu
         \,\pdmuPpz
\nl {}&-&
         \frac{1}{192}
         \,\frac{g^3\,M^2}{\ctW\,\pi^2}
         \,t^2_m\,t_1
         \,\Bigl[ 3\,t^2_m\,\mrA_0 - \lpar 1 - t_m \rpar^2 \Bigr]
         \,\Ppz
         \,\PZmu
         \,\pdmuPh
\nl {}&+&
         \frac{1}{192}
         \,\frac{g^3\,M^2}{\ctW\,\pi^2}
         \,t^2_m\,t_1
         \,\Bigl[ \lpar 1 - t_m \rpar\,\lpar 17 + t_m \rpar + 3\,
                  \lpar 9 - 9\,t_m + t^2_m \rpar\,\mrA_0 \Bigr]
         \,\Ph
         \,\PZmu
         \,\pdmuPpz
\nl {}&-&
         \frac{1}{512}
         \,\frac{g^4\,M^2}{\pi^2}
         \,\frac{t^2_m}{t_1}
         \,\lpar 2\,\PWpdmu\,\PWmdmu + \frac{1}{\ctWs}\,\PZdmus \rpar
         \,\PKz
         \,\beta^{(0)}_1
\nl {}&+&
         \frac{1}{512}
         \,\frac{g^4\,M^2}{\pi^2}
         \,\frac{t^2_m}{t_1}
         \,\lpar  2\,\beta^{(0)}_2 - 3\,\beta^{(0)}_1  \rpar
         \,\lpar 2\,\PWpdmu\,\PWmdmu + \frac{1}{\ctWs}\,\PZdmus \rpar
         \,\Ph^2
\nl {}&-&
         \frac{1}{1024}
         \,\frac{g^4\,M^2}{\pi^2}
         \,\frac{t^2_m}{t_1}
         \,x_{\Ph}
         \,\PKzs
         \,\beta^{(0)}_1
       +
         \frac{1}{1024}
         \,\frac{g^4\,M^2}{\pi^2}
         \,\frac{t^2_m}{t_1}
         \,x_{\Ph}
         \,\lpar  2\,\beta^{(0)}_2 - 5\,\beta^{(0)}_1  \rpar
         \,\Ph^4
\nl {}&+&
         \frac{1}{512}
         \,\frac{g^4\,M^2}{\pi^2}
         \,\frac{t^2_m}{t_1}
         \,x_{\Ph}
         \,\lpar  \beta^{(0}_2 - 3\,\beta^{(0)}_1  \rpar
         \,\PKz
         \,\Ph^2
\nl {}&-&
         \frac{1}{512}
         \,\frac{g^4\,M^2}{\pi^2}
         \,t^2_m
         \,\PKhs
         \,\beta^{(1)}_1
       +
         \frac{1}{512}
         \,\frac{g^4\,M^2}{\pi^2}
         \,t^2_m
         \,\PKhs
         \,\mrT_1
\nl {}&+&
         \frac{1}{1024}
         \,\frac{g^4\,M^2}{\pi^2}
         \,t^2_m
         \,\Bigl[ x^2_{\Ph}\,\mrA_0 - 32\,\lpar 1 - t_m \rpar\,
                 \lpar 1 + 3\,t_m - 5\,t^2_m \rpar\,t_m\,t^2_1 - 
               8\,\lpar 6 + 3\,t_m - 5\,t^2_m \rpar\,t_1\,x_{\Ph} 
\nl {}&-& 
               8\,\lpar 21 - 14\,t_m - 38\,t^2_m + 35\,t^3_m \rpar\,t_m\,t^2_1\,\mrA_0 -   
               2\,\lpar 36 + 17\,t_m - 14\,t^2_m \rpar\,t_1\,x_{\Ph}\,\mrA_0
       \Bigr]
         \,\PKz
         \,\Ph^2
\nl {}&+&
         \frac{1}{6144}
         \,\frac{g^4\,M^2}{\pi^2}
         \,t^2_m
         \,\Bigl[ 3\,x^2_{\Ph}\,\mrA_0 + 192\,\lpar 1 - t_m \rpar\,t^3_m\,t^2_1 - 
                 4\,\lpar 1 + 40\,t_m - 29\,t^2_m \rpar\,t_1\,x_{\Ph} + 24\,
                   \lpar 13 - 14\,t_m \rpar\,t^3_m\,t^2_1\,\mrA_0 
\nl {}&-& 
                 6\,\lpar 39 - 23\,t_m \rpar\,t_m\,t_1\,x_{\Ph}\,\mrA_0 \Bigr]
         \,\PKzs
\nl {}&+&
         \frac{1}{6144}
         \,\frac{g^4\,M^2}{\pi^2}
         \,t^2_m
         \,\Bigl[ 3\,x^2_{\Ph}\,\mrA_0 - 192\,\lpar 1 - t_m \rpar\,
                 \lpar 1 + 3\,t_m - 4\,t^2_m \rpar\,t_m\,t^2_1 - 
             168\,\lpar 6 - 4\,t_m - 9\,t^2_m + 8\,t^3_m \rpar\,t_m\,t^2_1\,\mrA_0 
\nl {}&-& 
               4\,\lpar 71 - 4\,t_m - 31\,t^2_m \rpar\,t_1\,x_{\Ph} - 
               6\,\lpar 72 - 5\,t_m - 5\,t^2_m \rpar\,t_1\,x_{\Ph}\,\mrA_0 \Bigr]
         \,\Ph^4
\nl {}&-&
         \frac{1}{256}
         \,\frac{g^4\,M^2}{\pi^2}
         \,t^2_m\,t_1
         \,\Bigl[ 2\,\lpar 3 - t_m \rpar + \lpar 9 - 2\,t_m + 2\,t^2_m \rpar\,\mrA_0 \Bigr]
         \,\lpar 2\,\PWpdmu\,\PWmdmu + \frac{1}{\ctWs}\,\PZdmus \rpar
         \,\PKz
\nl {}&-&
         \frac{1}{768}
         \,\frac{g^4\,M^2}{\pi^2}
         \,t^2_m\,t_1
         \,\Bigl[ 2\,\lpar 26 - 19\,t_m - t^2_m \rpar + 
                 3\,\lpar 27 - 20\,t_m + 4\,t^2_m \rpar\,\mrA_0 \Bigr]
         \,\lpar 2\,\PWpdmu\,\PWmdmu + \frac{1}{\ctWs}\,\PZdmus \rpar
         \,\Ph^2
\nl {}&+&
         \frac{3}{32}
         \,\frac{g^4\,M^2}{\pi^2}
         \,t^3_m\,t^2_1
         \,\Bigl[ \lpar 1 - t_m \rpar^2 + \lpar 1 - t_m \rpar^2\,\mrA_0 \Bigr]
         \,\PKh
         \,\Ph^2
\nl {}&+&
         \frac{1}{256}
         \,\frac{g^4\,M^2}{\pi^2}
         \,t^3_m
         \,\lpar  2\,\beta^{(0}_2 - 3\,\beta^{(0)}_1  \rpar
         \,\PKh
         \,\Ph^2
       -
         \frac{1}{512}
         \,\frac{g^4\,M^2}{\pi^2}
         \,t^4_m
         \,\lpar  5\,\beta^{(0}_2 - 8\,\beta^{(0)}_1  \rpar
         \,\Ph^4
\nl {}&-&
         \frac{1}{256}
         \,\frac{g^4\,M^2}{\pi^2}
         \,t^4_m
         \,\lpar  3\,\beta^{(0}_2 - 5\,\beta^{(0)}_1  \rpar
         \,\PKz
         \,\Ph^2
       -
         \frac{1}{512}
         \,\frac{g^4\,M^2}{\pi^2}
         \,t^4_m
         \,\lpar  \beta^{(0}_2 - 2\,\beta^{(0)}_1  \rpar
         \,\PKzs
\nl {}&-&
         \frac{1}{4}
         \,\frac{i\,g^2\,\stWs}{\ctW}
         \,M^2\,t^2_m
         \,\lpar \Ppp\,\PWmmu - \Ppm\,\PWpmu \rpar
         \,\Ph
         \,\PZmu
\nl {}&-&
         \frac{1}{128}
         \,\frac{i\,g^3\,M^2}{\pi^2}
         \,t^2_m\,x_{\Ph}\,
         \lpar x_{\PQd}\,\PAQd\,\gamma^5\,\PQd + x_{\PQu}\,\PAQu\,\gamma^5\,\PQu \rpar
         \,\Ppz
         \,\mrA_0
\nl {}&-&
         \frac{1}{128}
         \,\frac{i\,g^3\,M^2}{\srt\,\pi^2}
         \,t^2_m\,x_{\Ph}\,x_{\PQd}
         \,\lpar \PAQu\,\gamma_{-}\,\Ppp\,\PQd - \PAQd\,\gamma_{+}\,\Ppm\,\PQu \rpar
         \,\mrA_0
\nl {}&+&
         \frac{1}{128}
         \,\frac{i\,g^3\,M^2}{\srt\,\pi^2}
         \,t^2_m\,x_{\Ph}\,x_{\PQu}
         \,\lpar \PAQu\,\gamma_{+}\,\Ppp\,\PQd - \PAQd\,\gamma_{-}\,\Ppm\,\PQu \rpar
         \,\mrA_0
\nl {}&-&
         \frac{1}{128}
         \,\frac{i\,g^4\,M^2}{\pi^2}
         \,\frac{t^2_m}{t_1}
         \,\frac{\stWs}{\ctW}
         \,\lpar \beta^{(0}_2 - \beta^{(0)}_1  \rpar
         \,\lpar \Ppp\,\PWmmu - \Ppm\,\PWpmu \rpar
         \,\Ph
         \,\PZmu
\nl {}&+&
         \frac{1}{192}
         \,\frac{i\,g^4\,M^2}{\pi^2}
         \,\frac{\stWs}{\ctW}
         \,t^2_m\,t_1
         \,\Bigl[ \lpar 1 - t_m \rpar\,\lpar 17 + t_m \rpar + 
               3\,\lpar 9 - 9\,t_m + t^2_m \rpar\,\mrA_0 \Bigr]
         \,\lpar \Ppp\,\PWmmu - \Ppm\,\PWpmu \rpar
         \,\Ph
         \,\PZmu
\nl {}&+&
         \frac{1}{4}
         \,\frac{g}{\ctW}
         \,M^2\,t^2_m
         \,\lpar  \Ppz\,\pdmuPh - \Ph\,\pdmuPpz  \rpar
         \,\PZmu
\nl {}&+&
         \frac{1}{8}
         \,g^2\,M^2\,t^2_m
         \,\lpar \PKh + \Ph^2 \rpar
         \,\lpar 2\,\PWpdmu\,\PWmdmu + \frac{1}{\ctWs}\,\PZdmus \rpar
\nl {}&+&
         \frac{1}{32}
         \,g^2\,M^2\,t^2_m\,x_{\Ph}
         \,\lpar \PKh + 6\,\Ph^2 \rpar
         \,\PKh
\nl {}&+&
         \frac{1}{4}
         \,g\,M^2\,t^2_m
         \,\Bigl[ \lpar \Ppp\,\PWmmu + \Ppm\,\PWpmu \rpar\,\pdmuPh - \lpar \PWpmu\,
      \PdmuPpm + \PWmmu\,\PdmuPpp \rpar\,\Ph \Bigr]
\nl {}&+&
         \frac{1}{4}
         \,g\,M^2\,t^2_m\,
         \lpar x_{\PQd}\,\PAQd\,\PQd + x_{\PQu}\,\PAQu\,\PQu \rpar
         \,\Ph
\eqa
\bqa
\blacktriangleright\;\Lag_{65} &=&
         \frac{1}{8}
         \,\frac{g^2}{\ctW}
         \,M\,t^2_m
         \,\lpar  2\,\Ppz\,\pdmuPKh - \PKh\,\pdmuPpz  \rpar
         \,\PZmu
\nl {}&-&
         \frac{1}{128}
         \,\frac{g^3\,M}{\pi^2}
         \,\frac{t^2_m}{t_1}
         \,\lpar  \beta^{(0}_2 - \beta^{(0)}_1  \rpar
         \,\pdmuPh
         \,\pdmuPKh
\nl {}&-&
         \frac{1}{192}
         \,\frac{g^3\,M}{\pi^2}
         \,t^2_m\,t_1
         \,\lpar 2 - 9\,\mrA_0 \rpar
         \,\lpar 1 - t_m \rpar
         \,\pdmuPh
         \,\pdmuPKh
\nl {}&+&
         \frac{1}{512}
         \,\frac{g^4\,M}{\pi^2}
         \,\frac{t^2_m}{t_1}
         \,\Bigl[ \lpar \PWpmu\,\PdmuPpm + \PWmmu\,\PdmuPpp \rpar\,\PKh
                   - 2\,\lpar \Ppp\,\PWmmu + \Ppm\,\PWpmu \rpar\,\pdmuPKh 
                    \Bigr]
         \,\beta^{(0)}_1
\nl {}&-&
         \frac{1}{512}
         \,\frac{g^4\,M}{\pi^2}
         \,\frac{t^2_m}{t_1}
         \,\beta^{(0)}_1
         \,\lpar x_{\PQu}\,\PAQu\,\PQu + x_{\PQd}\,\PAQd\,\PQd \rpar
         \,\PKh
\nl {}&+&
         \frac{1}{256}
         \,\frac{g^4\,M}{\pi^2}
         \,t^2_m\,t_1
         \,\Bigl[ 6 + \lpar 9 + t^2_m \rpar\,\mrA_0 \Bigr]
         \,\Bigl[ \lpar \PWpmu\,\PdmuPpm + \PWmmu\,\PdmuPpp \rpar\,\PKh 
                   - 2\,\lpar \Ppp\,\PWmmu + \Ppm\,\PWpmu \rpar\,\pdmuPKh 
                    \Bigr]
\nl {}&-&
         \frac{3}{256}
         \,\frac{g^4\,M}{\pi^2}
         \,t^2_m\,t_1
         \,\lpar 2 + 3\,\mrA_0 \rpar
         \,\lpar x_{\PQd}\,\PAQd\,\PQd + x_{\PQu}\,\PAQu\,\PQu \rpar
         \,\PKh
\nl {}&-&
         \frac{1}{512}
         \,\frac{g^4\,M}{\ctW\,\pi^2}
         \,\frac{t^2_m}{t_1}
         \,\lpar 2\,\Ppz\,\pdmuPKh - \PKh\,\pdmuPpz  \rpar
         \,\PZmu
         \,\beta^{(0)}_1
\nl {}&-&
         \frac{1}{256}
         \,\frac{g^4\,M}{\ctW\,\pi^2}
         \,t^2_m\,t_1
         \,\Bigl[ 6 + \lpar 9 + t^2_m \rpar\,\mrA_0 \Bigr]
         \,\lpar 2\,\Ppz\,\pdmuPKh - \PKh\,\pdmuPpz  \rpar
         \,\PZmu
\nl {}&-&
         \frac{1}{1024}
         \,\frac{g^5\,M}{\pi^2}
         \,\frac{t^2_m}{t_1}
         \,\lpar 2\,\PWpdmu\,\PWmdmu + \frac{1}{\ctWs}\,\PZdmus \rpar
         \,\PKh
         \,\Ph
         \,\beta^{(0)}_1
\nl {}&-&
         \frac{1}{2048}
         \,\frac{g^5\,M}{\pi^2}
         \,\frac{t^2_m}{t_1}
         \,\lpar x_{\Ph} - 4\,t^2_m\,t_1 \rpar
         \,\Ph
         \,\PKhs
         \,\beta^{(0)}_1
\nl {}&-&
         \frac{1}{512}
         \,\frac{g^5\,M}{\pi^2}
         \,t^2_m\,t_1
         \,\Bigl[ 6 + \lpar 9 + t^2_m \rpar\,\mrA_0 \Bigr]
         \,\lpar 2\,\PWpdmu\,\PWmdmu + \frac{1}{\ctWs}\,\PZdmus \rpar
         \,\PKh
         \,\Ph
\nl {}&-&
         \frac{1}{1024}
         \,\frac{g^5\,M}{\pi^2}
         \,t^2_m\,t_1
         \,\Bigl[ 6\,x_{\Ph} + 4\,\lpar 1 - t_m \rpar\,
                 \lpar 5 + 4\,t_m - 4\,t^2_m \rpar\,t_m\,t_1 + 
                 \lpar 9 + 2\,t^2_m \rpar\,x_{\Ph}\,\mrA_0 
\nl {}&+& 
               2\,\lpar 15 - t_m - 26\,t^2_m + 14\,t^3_m \rpar\,t_m\,t_1\,\mrA_0 \Bigr]
         \,\Ph
         \,\PKhs
\nl {}&+&
         \frac{1}{1024}
         \,\frac{g^5\,M}{\pi^2}
         \,t^3_m
         \,\lpar  \beta^{(0}_2 - 3\,\beta^{(0)}_1  \rpar
         \,\Ph
         \,\PKhs
\nl {}&+&
         \frac{3}{128}
         \,\frac{g^5\,M}{\pi^2}
         \,t^3_m\,t^2_1
         \,\lpar 1 + \mrA_0 \rpar
         \,\lpar 1 - t_m \rpar
         \,\Ph
         \,\PKhs
\nl {}&-&
         \frac{1}{8}
         \,\frac{i\,g^3\,\stWs}{\ctW}
         \,M\,t^2_m
         \,\lpar \Ppp\,\PWmmu - \Ppm\,\PWpmu \rpar
         \,\PKh
         \,\PZmu
\nl {}&+&
         \frac{1}{512}
         \,\frac{i\,g^5\,M}{\pi^2}
         \,\frac{t^2_m}{t_1}
         \,\frac{\stWs}{\ctW}
         \,\lpar \Ppp\,\PWmmu - \Ppm\,\PWpmu \rpar
         \,\PKh
         \,\PZmu
         \,\beta^{(0)}_1
\nl {}&+&
         \frac{1}{256}
         \,\frac{i\,g^5\,M}{\pi^2}
         \,\frac{\stWs}{\ctW}
         \,t^2_m\,t_1
         \,\Bigl[ 6 + \lpar 9 + t^2_m \rpar\,\mrA_0 \Bigr]
         \,\lpar \Ppp\,\PWmmu - \Ppm\,\PWpmu \rpar
         \,\PKh
         \,\PZmu
\nl {}&-&
         \frac{1}{8}
         \,g^2\,M\,t^2_m
         \,\Bigl[ \lpar \PWpmu\,\PdmuPpm + \PWmmu\,\PdmuPpp \rpar\,\PKh 
                  - 2\,\lpar \Ppp\,\PWmmu + \Ppm\,\PWpmu \rpar\,\pdmuPKh 
                       \Bigr]
\nl {}&+&
         \frac{1}{8}
         \,g^2\,M\,t^2_m\,
         \lpar x_{\PQd}\,\PAQd\,\PQd + x_{\PQu}\,\PAQu\,\PQu \rpar
         \,\PKh
\nl {}&+&
         \frac{1}{16}
         \,g^3\,M\,t^2_m
         \,\lpar 2\,\PWpdmu\,\PWmdmu + \frac{1}{\ctWs}\,\PZdmus \rpar
         \,\PKh
         \,\Ph
       +
         \frac{1}{32}
         \,g^3\,M\,t^2_m\,x_{\Ph}
         \,\Ph
         \,\PKhs \spc
\nl
{}&+&
         \frac{3}{256}
         \,\frac{g^5\,\,M}{\pi^2}
         \,t^3_m\,t^2_1)
         \,(1 - t_m)
         \,\mrB_{00}
         \,\Ph
         \,\PKhs
\eqa
\bqa
\blacktriangleright\;\Lag_{66} &=&
       -
         \frac{1}{384}
         \,\frac{g^4}{\pi^2}
         \,t^2_m\,t_1
         \,\pdmuPKhs -
         \frac{1}{4096}
         \,\frac{g^6}{\pi^2}
         \,t^3_m
         \,\PKhc
         \,\beta^{(0)}_1 
\nl {}&-&
         \frac{1}{3072}
         \,\frac{g^6}{\pi^2}
         \,t^3_m\,t^2_1
         \,\lpar 5 + 9\,\mrA_0 \rpar
         \,\PKhc -
         \frac{1}{8}
         \,g^2\,t^2_m
         \,\pdmuPKhs \spp
\nl
{}&+&  \frac{3}{1024}
         \,\frac{g^6}{\pi^2}
         \,t^3_m\,t^2_1
         \,\mrB_{00}
         \,\PKhc
\eqa
\eei


%


 \clearpage
\bibliographystyle{elsarticle-num}
\bibliography{LExp}

\begin{thebibliography}{10}
\expandafter\ifx\csname url\endcsname\relax
  \def\url#1{\texttt{#1}}\fi
\expandafter\ifx\csname urlprefix\endcsname\relax\def\urlprefix{URL }\fi
\expandafter\ifx\csname href\endcsname\relax
  \def\href#1#2{#2} \def\path#1{#1}\fi

\bibitem{deBlas:2014mba}
J.~de~Blas, M.~Chala, M.~Perez-Victoria, J.~Santiago, {Observable Effects of
  General New Scalar Particles}, JHEP 04 (2015) 078.
\newblock \href {http://arxiv.org/abs/1412.8480} {\path{arXiv:1412.8480}},
  \href {http://dx.doi.org/10.1007/JHEP04(2015)078}
  {\path{doi:10.1007/JHEP04(2015)078}}.

\bibitem{Chiang:2015ura}
C.-W. Chiang, R.~Huo, {Standard Model Effective Field Theory: Integrating out a
  Generic Scalar}, JHEP 09 (2015) 152.
\newblock \href {http://arxiv.org/abs/1505.06334} {\path{arXiv:1505.06334}},
  \href {http://dx.doi.org/10.1007/JHEP09(2015)152}
  {\path{doi:10.1007/JHEP09(2015)152}}.

\bibitem{Brehmer:2015rna}
J.~Brehmer, A.~Freitas, D.~Lopez-Val, T.~Plehn, {Pushing Higgs Effective Theory
  to its Limits}\href {http://arxiv.org/abs/1510.03443}
  {\path{arXiv:1510.03443}}.

\bibitem{Biekotter:2016ecg}
A.~Biekötter, J.~Brehmer, T.~Plehn, {Pushing Higgs Effective Theory over the
  Edge}\href {http://arxiv.org/abs/1602.05202} {\path{arXiv:1602.05202}}.

\bibitem{delAguila:2016zcb}
F.~del Aguila, Z.~Kunszt, J.~Santiago, {One-loop effective lagrangians after
  matching}\href {http://arxiv.org/abs/1602.00126} {\path{arXiv:1602.00126}}.

\bibitem{Gaillard:1985uh}
M.~K. Gaillard, {The Effective One Loop Lagrangian With Derivative Couplings},
  Nucl. Phys. B268 (1986) 669.
\newblock \href {http://dx.doi.org/10.1016/0550-3213(86)90264-6}
  {\path{doi:10.1016/0550-3213(86)90264-6}}.

\bibitem{Cheyette:1987qz}
O.~Cheyette, {Effective Action for the Standard Model With Large Higgs Mass},
  Nucl. Phys. B297 (1988) 183.
\newblock \href {http://dx.doi.org/10.1016/0550-3213(88)90205-2}
  {\path{doi:10.1016/0550-3213(88)90205-2}}.

\bibitem{Henning:2014wua}
B.~Henning, X.~Lu, H.~Murayama, {How to use the Standard Model effective field
  theory}, JHEP 01 (2016) 023.
\newblock \href {http://arxiv.org/abs/1412.1837} {\path{arXiv:1412.1837}},
  \href {http://dx.doi.org/10.1007/JHEP01(2016)023}
  {\path{doi:10.1007/JHEP01(2016)023}}.

\bibitem{Drozd:2015rsp}
A.~Drozd, J.~Ellis, J.~Quevillon, T.~You, {The Universal One-Loop Effective
  Action}\href {http://arxiv.org/abs/1512.03003} {\path{arXiv:1512.03003}}.

\bibitem{SekharChivukula:2007gi}
R.~S. Chivukula, N.~D. Christensen, E.~H. Simmons, {Low-energy effective
  theory, unitarity, and non-decoupling behavior in a model with heavy
  Higgs-triplet fields}, Phys. Rev. D77 (2008) 035001.
\newblock \href {http://arxiv.org/abs/0712.0546} {\path{arXiv:0712.0546}},
  \href {http://dx.doi.org/10.1103/PhysRevD.77.035001}
  {\path{doi:10.1103/PhysRevD.77.035001}}.

\bibitem{Chen:2008jg}
M.-C. Chen, S.~Dawson, C.~B. Jackson, {Higgs Triplets, Decoupling, and
  Precision Measurements}, Phys. Rev. D78 (2008) 093001.
\newblock \href {http://arxiv.org/abs/0809.4185} {\path{arXiv:0809.4185}},
  \href {http://dx.doi.org/10.1103/PhysRevD.78.093001}
  {\path{doi:10.1103/PhysRevD.78.093001}}.

\bibitem{Georgi:1985nv}
H.~Georgi, M.~Machacek, {Doubly charged Higgs bosons}, Nucl. Phys. B262 (1985)
  463.
\newblock \href {http://dx.doi.org/10.1016/0550-3213(85)90325-6}
  {\path{doi:10.1016/0550-3213(85)90325-6}}.

\bibitem{Einhorn:1981cy}
M.~B. Einhorn, D.~R.~T. Jones, M.~J.~G. Veltman, {Heavy Particles and the rho
  Parameter in the Standard Model}, Nucl. Phys. B191 (1981) 146.
\newblock \href {http://dx.doi.org/10.1016/0550-3213(81)90292-3}
  {\path{doi:10.1016/0550-3213(81)90292-3}}.

\bibitem{Low:2012rj}
I.~Low, J.~Lykken, G.~Shaughnessy, {Have We Observed the Higgs (Imposter)?},
  Phys. Rev. D86 (2012) 093012.
\newblock \href {http://arxiv.org/abs/1207.1093} {\path{arXiv:1207.1093}},
  \href {http://dx.doi.org/10.1103/PhysRevD.86.093012}
  {\path{doi:10.1103/PhysRevD.86.093012}}.

\bibitem{Kanemura:2014bqa}
S.~Kanemura, K.~Tsumura, K.~Yagyu, H.~Yokoya, {Fingerprinting nonminimal Higgs
  sectors}, Phys. Rev. D90 (2014) 075001.
\newblock \href {http://arxiv.org/abs/1406.3294} {\path{arXiv:1406.3294}},
  \href {http://dx.doi.org/10.1103/PhysRevD.90.075001}
  {\path{doi:10.1103/PhysRevD.90.075001}}.

\bibitem{Dittmaier:1995cr}
S.~Dittmaier, C.~Grosse-Knetter, {Deriving nondecoupling effects of heavy
  fields from the path integral: A Heavy Higgs field in an SU(2) gauge theory},
  Phys. Rev. D52 (1995) 7276--7293.
\newblock \href {http://arxiv.org/abs/hep-ph/9501285}
  {\path{arXiv:hep-ph/9501285}}, \href
  {http://dx.doi.org/10.1103/PhysRevD.52.7276}
  {\path{doi:10.1103/PhysRevD.52.7276}}.

\bibitem{Silveira:1985rk}
V.~Silveira, A.~Zee, {Scalar phantoms}, Phys. Lett. B161 (1985) 136.
\newblock \href {http://dx.doi.org/10.1016/0370-2693(85)90624-0}
  {\path{doi:10.1016/0370-2693(85)90624-0}}.

\bibitem{Schabinger:2005ei}
R.~Schabinger, J.~D. Wells, {A Minimal spontaneously broken hidden sector and
  its impact on Higgs boson physics at the large hadron collider}, Phys. Rev.
  D72 (2005) 093007.
\newblock \href {http://arxiv.org/abs/hep-ph/0509209}
  {\path{arXiv:hep-ph/0509209}}, \href
  {http://dx.doi.org/10.1103/PhysRevD.72.093007}
  {\path{doi:10.1103/PhysRevD.72.093007}}.

\bibitem{Pruna:2013bma}
G.~M. Pruna, T.~Robens, {Higgs singlet extension parameter space in the light
  of the LHC discovery}, Phys. Rev. D88~(11) (2013) 115012.
\newblock \href {http://arxiv.org/abs/1303.1150} {\path{arXiv:1303.1150}},
  \href {http://dx.doi.org/10.1103/PhysRevD.88.115012}
  {\path{doi:10.1103/PhysRevD.88.115012}}.

\bibitem{Robens:2015gla}
T.~Robens, T.~Stefaniak, {Status of the Higgs Singlet Extension of the Standard
  Model after LHC Run 1}, Eur. Phys. J. C75 (2015) 104.
\newblock \href {http://arxiv.org/abs/1501.02234} {\path{arXiv:1501.02234}},
  \href {http://dx.doi.org/10.1140/epjc/s10052-015-3323-y}
  {\path{doi:10.1140/epjc/s10052-015-3323-y}}.

\bibitem{Robens:2016xkb}
T.~Robens, T.~Stefaniak, {LHC Benchmark Scenarios for the Real Higgs Singlet
  Extension of the Standard Model}\href {http://arxiv.org/abs/1601.07880}
  {\path{arXiv:1601.07880}}.

\bibitem{Gorbahn:2015gxa}
M.~Gorbahn, J.~M. No, V.~Sanz, {Benchmarks for Higgs Effective Theory: Extended
  Higgs Sectors}, JHEP 10 (2015) 036.
\newblock \href {http://arxiv.org/abs/1502.07352} {\path{arXiv:1502.07352}},
  \href {http://dx.doi.org/10.1007/JHEP10(2015)036}
  {\path{doi:10.1007/JHEP10(2015)036}}.

\bibitem{Actis:2006ra}
S.~Actis, A.~Ferroglia, M.~Passera, G.~Passarino, {Two-Loop Renormalization in
  the Standard Model. Part I: Prolegomena}, Nucl. Phys. B777 (2007) 1--34.
\newblock \href {http://arxiv.org/abs/hep-ph/0612122}
  {\path{arXiv:hep-ph/0612122}}, \href
  {http://dx.doi.org/10.1016/j.nuclphysb.2007.04.021}
  {\path{doi:10.1016/j.nuclphysb.2007.04.021}}.

\bibitem{Kallosh:1972ap}
R.~E. Kallosh, I.~V. Tyutin, {The Equivalence theorem and gauge invariance in
  renormalizable theories}, Yad. Fiz. 17 (1973) 190--209, [Sov. J. Nucl.
  Phys.17,98(1973)].

\bibitem{Arzt:1993gz}
C.~Arzt, {Reduced effective Lagrangians}, Phys. Lett. B342 (1995) 189--195.
\newblock \href {http://arxiv.org/abs/hep-ph/9304230}
  {\path{arXiv:hep-ph/9304230}}, \href
  {http://dx.doi.org/10.1016/0370-2693(94)01419-D}
  {\path{doi:10.1016/0370-2693(94)01419-D}}.

\bibitem{Tyutin:2000ht}
I.~V. Tyutin, {Once again on the equivalence theorem}, Phys. Atom. Nucl. 65
  (2002) 194--202, [Yad. Fiz.65,201(2002)].
\newblock \href {http://arxiv.org/abs/hep-th/0001050}
  {\path{arXiv:hep-th/0001050}}, \href {http://dx.doi.org/10.1134/1.1446571}
  {\path{doi:10.1134/1.1446571}}.

\bibitem{Gavela:2016bzc}
B.~M. Gavela, E.~E. Jenkins, A.~V. Manohar, L.~Merlo, {Analysis of General
  Power Counting Rules in Effective Field Theory}\href
  {http://arxiv.org/abs/1601.07551} {\path{arXiv:1601.07551}}.

\bibitem{Grzadkowski:2010es}
B.~Grzadkowski, M.~Iskrzynski, M.~Misiak, J.~Rosiek, {Dimension-Six Terms in
  the Standard Model Lagrangian}, JHEP 1010 (2010) 085.
\newblock \href {http://arxiv.org/abs/1008.4884} {\path{arXiv:1008.4884}},
  \href {http://dx.doi.org/10.1007/JHEP10(2010)085}
  {\path{doi:10.1007/JHEP10(2010)085}}.

\bibitem{'tHooft:1972ue}
G.~'t~Hooft, M.~J.~G. Veltman, {Combinatorics of gauge fields}, Nucl. Phys. B50
  (1972) 318--353.
\newblock \href {http://dx.doi.org/10.1016/S0550-3213(72)80021-X}
  {\path{doi:10.1016/S0550-3213(72)80021-X}}.

\bibitem{Alonso:2015fsp}
R.~Alonso, E.~E. Jenkins, A.~V. Manohar, {A Geometric Formulation of Higgs
  Effective Field Theory: Measuring the Curvature of Scalar Field Space}\href
  {http://arxiv.org/abs/1511.00724} {\path{arXiv:1511.00724}}.

\bibitem{Gavela:2014uta}
M.~B. Gavela, K.~Kanshin, P.~A.~N. Machado, S.~Saa, {On the renormalization of
  the electroweak chiral Lagrangian with a Higgs}, JHEP 03 (2015) 043.
\newblock \href {http://arxiv.org/abs/1409.1571} {\path{arXiv:1409.1571}},
  \href {http://dx.doi.org/10.1007/JHEP03(2015)043}
  {\path{doi:10.1007/JHEP03(2015)043}}.

\bibitem{Buchalla:2015qju}
G.~Buchalla, O.~Cata, A.~Celis, C.~Krause, {Fitting Higgs Data with Nonlinear
  Effective Theory}\href {http://arxiv.org/abs/1511.00988}
  {\path{arXiv:1511.00988}}.

\bibitem{Branco:2011iw}
G.~C. Branco, P.~M. Ferreira, L.~Lavoura, M.~N. Rebelo, M.~Sher, J.~P. Silva,
  {Theory and phenomenology of two-Higgs-doublet models}, Phys. Rept. 516
  (2012) 1--102.
\newblock \href {http://arxiv.org/abs/1106.0034} {\path{arXiv:1106.0034}},
  \href {http://dx.doi.org/10.1016/j.physrep.2012.02.002}
  {\path{doi:10.1016/j.physrep.2012.02.002}}.

\bibitem{Yagyu:2012qp}
K.~Yagyu,
  \href{http://inspirehep.net/record/1097017/files/arXiv:1204.0424.pdf}{{Studies
  on Extended Higgs Sectors as a Probe of New Physics Beyond the Standard
  Model}}, Ph.D. thesis, Toyama U. (2012).
\newblock \href {http://arxiv.org/abs/1204.0424} {\path{arXiv:1204.0424}}.
\newline\urlprefix\url{http://inspirehep.net/record/1097017/files/arXiv:1204.0424.pdf}

\bibitem{Bhattacharyya:2015nca}
G.~Bhattacharyya, D.~Das, {Scalar sector of Two-Higgs-Doublet models: A
  mini-review}\href {http://arxiv.org/abs/1507.06424}
  {\path{arXiv:1507.06424}}.

\bibitem{Gunion:2002zf}
J.~F. Gunion, H.~E. Haber, {The CP conserving two Higgs doublet model: The
  Approach to the decoupling limit}, Phys. Rev. D67 (2003) 075019.
\newblock \href {http://arxiv.org/abs/hep-ph/0207010}
  {\path{arXiv:hep-ph/0207010}}, \href
  {http://dx.doi.org/10.1103/PhysRevD.67.075019}
  {\path{doi:10.1103/PhysRevD.67.075019}}.

\bibitem{Carena:2013ooa}
M.~Carena, I.~Low, N.~R. Shah, C.~E.~M. Wagner, {Impersonating the Standard
  Model Higgs Boson: Alignment without Decoupling}, JHEP 04 (2014) 015.
\newblock \href {http://arxiv.org/abs/1310.2248} {\path{arXiv:1310.2248}},
  \href {http://dx.doi.org/10.1007/JHEP04(2014)015}
  {\path{doi:10.1007/JHEP04(2014)015}}.

\end{thebibliography}

\end{document}